\documentclass[conference]{IEEEtran}
\IEEEoverridecommandlockouts

\usepackage{cite}
\usepackage{amsmath,amssymb,amsfonts}
\usepackage{algorithmic}
\usepackage{graphicx}
\usepackage{textcomp}
\usepackage{xcolor}
\usepackage{my_sty}
\usepackage{tabularx} 

\usepackage{titlesec}
\titlespacing*{\section}{0pt}{1ex plus 1ex minus .2ex}{1ex plus .2ex}

\newif\ififip  
\ifiptrue   
\newcommand{\hideforifip}[1]{\ififip\else#1\fi}

\newcommand{\showforifip}[1]{\ififip#1\else\fi}

\newcommand{\ignore}[1]{}

\usepackage[letterpaper, left=1.01in, right=1.01in, top=0.75in, bottom=1.05in]{geometry}
\begin{document}

\title{LAPRAD: LLM-Assisted PRotocol Attack Discovery\\
}

\showforifip{
\author{
{\rm R. Can Aygun}\\
rcaygun@cs.ucla.edu\\
UCLA
\and
{\rm Yehuda Afek }\\
afek@tauex.tau.ac.il\\
Tel-Aviv University
\and
{\rm Anat Bremler-Barr}\\
anatbr@tauex.tau.ac.il\\
Tel-Aviv University
\and
{\rm Leonard Kleinrock}\\
lk@cs.ucla.edu \\
UCLA
} 

}
\maketitle
\begin{abstract}
With the goal of improving the security of Internet protocols, we seek faster, semi-automatic methods to discover new vulnerabilities in protocols such as DNS, BGP, and others. To this end, we introduce the LLM-Assisted Protocol Attack Discovery (LAPRAD) methodology, enabling security researchers with some DNS knowledge to efficiently uncover vulnerabilities that would otherwise be hard to detect.

LAPRAD follows a three-stage process.
In the first, we consult an LLM (GPT‑o1) that has been trained on a broad corpus of DNS-related sources and previous DDoS attacks to identify potential exploits.
In the second stage, a different LLM automatically constructs the corresponding attack configurations using the ReACT approach implemented via LangChain (DNS zone file generation).
Finally, in the third stage, we validate the attack’s functionality and effectiveness.

Using LAPRAD, we uncovered three new DDoS attacks on the DNS protocol and rediscovered two recently reported ones that were not included in the LLM’s training data. The first new attack employs a bait-and-switch technique to trick resolvers into caching large, bogus DNSSEC RRSIGs, reducing their serving capacity to as little as 6\%. The second exploits large DNSSEC encryption algorithms (RSA‑4096) with multiple keys, thereby bypassing a recently implemented default RRSet limit. The third leverages ANY-type responses to produce a similar effect.
These variations of a cache-flushing DDoS attack, called SigCacheFlush, circumvent existing patches, severely degrade resolver query capacity, and impact the latest versions of major DNS resolver implementations.

\ignore{
With the goal of making Internet protocols more secure, we seek faster, semi-automatic methods to discover new vulnerabilities in protocols such as DNS, BGP, and others.
To this end, this paper introduces an LLM-Assisted Protocol Attack Discovery (LAPRAD) methodology, enabling security researchers familiar with the DNS to efficiently uncover DNS vulnerabilities that would otherwise be very difficult to spot.
LAPRAD is a three-stage methodology. In the first stage, we interact with an LLM (GPT-o1) that has a large corpus of training data consisting of DNS related sources,
\ignore{including RFCs, public security discussions,} 
previous DDoS attacks, to identify potential attack ideas\ignore{, with the LLM providing reasoning and citing specific RFC's}. In the second phase, a different LLM is used to construct a corresponding attack configuration (including zone file generation), and in the third stage, the attack is tested for its functionality and effectiveness.
}

\ignore{
Using LAPRAD, we uncovered three new DNS protocol vulnerabilities and rediscovered two recently reported vulnerabilities that were not part of the LLM's training data.
The first newly discovered attack uses a bait-and-switch technique to trick the resolver into caching huge bogus DNSSEC RRSIGs.
This attack cripples a resolver to serve at most $6\%$ of its full capacity.
The second leverages large DNSSEC encryption algorithms (RSA-4096) with multiple keys, rendering a recently implemented RRSet default limit ineffective.
The third utilizes ANY-type responses to achieve a similar effect.
These new vulnerabilities, which are variants of the cache-flushing DDoS attack, bypass existing patches, severely degrade resolver query capacity, and impact the latest versions of major DNS resolver implementations.
}

\ignore{
These over-sized DNSSEC responses, employed in all three attack variants, evict benign domains from the resolver cache, forcing frequent re-queries to authoritative servers.
During the tests, the bait-and-switch variant reduced benign traffic throughput by up to 94\% and achieved three times more cache flushing capacity than previous CacheFlushing attacks at the same attack query rate.

In addition to the three new vulnerabilities we successfully rediscovered with the assistance of LAPRAD two recent DNS attacks, KeyTrap and CacheFlush, which were not part of the used LLM's training data.
}

\ignore{
As an example of its use, we show it effectively uncovers new DNS protocol vulnerabilities by analyzing the logic and rules outlined in DNS RFCs.\\rca{We demonstrate its effectiveness in uncovering new DNS protocol vulnerabilities by leveraging its training on extensive DNS-related documents, such as RFCs and public sources.} In this way discovered three new variants of the Cache-Flushing DDoS vulnerability based on the DNSSEC extension.
}

\ignore{
As part of the LAPRAD methodology, we employed a multi-step strategy to prompt a second LLM to generate detailed attack configurations and validate its reasoning using specific RFC references.
We also used an LLM to identify effective mitigations for newly discovered vulnerabilities.
Lastly, in addition we successfully rediscovered two recent DNS attacks, KeyTrap and CacheFlush, which were not part of the LLM's training data.
}

\end{abstract}


\section{Introduction}

Internet protocols are inherently complex and vulnerable, requiring researchers to quickly detect and fix weaknesses. Several studies, such as GRoot\cite{kakarla_groot_2020}, and the work by Siu et al.\cite{liu_formal_2023}, have explored formal modeling approaches for DNS, aiming to systematically analyze the protocol and uncover vulnerabilities. However, these approaches require substantial manual effort to model the protocol and result in only a partial protocol model, limiting the verification of protocol logic, as is the case in the DNS protocol which is complex with 297 RFCs\cite{noauthor_powerdns_nodate}.

\ignore{
Internet network protocols are complex, distributed, and constantly evolving—an environment ripe with vulnerabilities that attackers can, and do exploit for malicious purposes. To stay ahead, security researchers must rapidly discover and patch these weaknesses.

{Several studies, such as GRoot\cite{kakarla_groot_2020}, SCALE\cite{kakarla_scale_2022}, and the work by Siu et al.\cite{liu_formal_2023}, have explored formal modeling approaches for DNS, aiming to systematically analyze the protocol and uncover vulnerabilities. However, these approaches require substantial manual effort to model the protocol and result in only a partial protocol model, limiting the verification of protocol logic, as the DNS protocol is highly complex with 297 approved RFCs\cite{noauthor_powerdns_nodate}.}
}

\ignore{
In this paper we introduce an LLM-Assisted Protocol Attack Discovery (LAPRAD) methodology, empowering DNS-focused security researchers to rapidly and efficiently identify otherwise elusive vulnerabilities.

DNS is one of the most critical and complex components of the internet, and it is frequently targeted by attackers. Kaminsky’s cache poisoning vulnerability redirected users to malicious websites\cite{kaminsky_black_nodate}. TsuNAME leverages cyclic dependencies in DNS records to cause recursive resolvers to loop, overloading servers \cite{moura_tsuname_2021}. KeyTrap exploited DNSSEC protocol flaws to overwhelm resolvers with computational complexity\cite{heftrig_harder_2024}. CacheFlushing attack fills benign resolver caches\cite{afek2024flushing} with many bogus NS responses. These vulnerabilities could have disrupted internet services and required immediate patching.

Many vulnerabilities have been discovered years after their introduction into protocol specifications and implementations. The complexity of DNS, with its 297 approved RFCs\cite{noauthor_dns_nodate} and diverse implementations across vendors like BIND \cite{noauthor_bind_nodate}, Unbound \cite{noauthor_nlnet_nodate}, and Knot \cite{noauthor_knot_nodate}, increases the likelihood of exposure to attacks and makes investigating them particularly challenging.
This highlights the need for faster and more efficient techniques to identify protocol weaknesses in a timely manner.
}

In recent years, large language models (LLMs) have shown success in tasks like natural language understanding \cite{devlin_bert_2018}, text generation \cite{brown_language_2020}, and code generation \cite{chen_evaluating_2021}.



In this paper we harness LLMs technology and prompt engineering to develop LAPRAD, a methodology to assist in the fast and efficient discovery of vulnerabilities in Internet protocols. 
We show how LAPRAD enables a security researcher familiar with the DNS protocol to identify potential DNS protocol DDoS vulnerabilities, and test these ideas in a controlled testing environment in an automated fashion.
It follows three key steps: (1) the researcher provides an example attack to the LLM, which generates similar attack ideas and refines them through iterations.
(2) another LLM automatically generates a suitable zone file for the discovered attack idea using the ReACT approach; and
(3) the researcher tests the attack using the generated zone file in a real testing environment.

\ignore{This methodology lowers the barrier to discovering new DoS vulnerabilities in the DNS protocol, a task that traditionally requires deep expertise. LAPRAD, however, enables security researchers familiar with the DNS protocol to effectively explore and develop attack scenarios. }

This methodology leverages LLMs' extensive knowledge of DNS protocol features and security warnings from RFCs and public sources (e.g., DNS-OARC, IETF DNSOP mailing lists), offering a faster alternative to the labor-intensive process of manually analyzing protocol documents and security forums. 
Moreover, LLMs retain information from remote, hard-to-find sources that may no longer be available, such as public mailing lists and conference websites. Thus serving as a valuable resource for future vulnerability research.

\ignore{
In addition, we use LLM agents to automate the generation of attack-related DNS zone files, a process known to be error-prone. We also set up a controlled testing environment managed by a single script, enabling faster testing of attack ideas developed with LLM assistance.}

We use LAPRAD (our main contribution), to make the following additional contributions:
\begin{itemize} 
    \item Identified three distinct new variants of the CacheFlushing vulnerability based on DNSSEC.
    These variants bypass recently issued CacheFlushing patches\cite{CVE-2024-1737} affecting the latest versions of major DNS resolver implementations (BIND 9.18.31\cite{noauthor_bind_nodate}, Unbound 1.22.0\cite{noauthor_nlnet_nodate}, and Knot 5.7.4\cite{noauthor_knot_nodate}), significantly degrading their throughput. 
    \item Automatically generate attack zone files (an error-prone process), along with a high-level execution guide to comprehensively illustrate the discovered vulnerabilities (Section~\ref{subsubsec:gen-attack-configs}).
    \item Assisting in pinpointing exact RFC sections and sentences that represent the root cause of the discovered vulnerabilities, while providing clear reasoning explaining why these elements contribute to the vulnerabilities (Section~\ref{subsubsec:source-of-info}).
    \item Re-discovering two recent DNS attacks, KeyTrap \cite{heftrig_harder_2024} and CacheFlush \cite{afek2024flushing}, without these attacks being part of the LLM’s training data. 
\end{itemize}

\hideforifip{The previously discovered low-rate CacheFlush attack relied on packing a large number of resource records (up to 2,000 NS records) into a single packet, generating a 65kB response for a single malicious query. It was reported that an attacker generating just 1Kqps(queries per second) could flush a resolver with a 100MB cache , reducing the resolver's throughput by 80\%. To mitigate this attack, recent patches have limited the maximum number of records per RRSet to 100, making this attack infeasible with standard DNS RRSets.}

\hideforifip{The previous CacheFlush\cite{afek2024flushing} attack exploited large resource record sets, packing up to 2,000 NS records into a single packet and generating a 65kB response per malicious query. An attacker sending 1K qps (queries per second) could flush a 100MB resolver cache, reducing throughput by 80\%. Recent patches mitigate this by capping RRSets at 100 records, rendering the attack infeasible with standard DNS RRSets.}
\hideforifip{
The \textit{SigCacheFlush} attack we discovered using LAPRAD bypasses these patches with three distinct techniques: (1) Bait \&Switch technique using a single large signature; (2) multiple large-sized keys; and (3) ANY-type responses. This attack enables the insertion of even larger amounts of bogus data into the resolver’s cache with a single query. A malicious reply generates a ~65KB DNSSEC response, which is repeated three times for additional queries related to DS, NS, and DNSKEY records for the domain.
This inserts ~195KB of bogus data into the resolver’s cache per malicious query.}


\ignore{
We set up a test environment to assess the attack's impact on the latest versions of BIND, Unbound, and Knot resolvers. Tests were run on BIND 9.18.31\cite{noauthor_bind_nodate}, Unbound 1.22.0\cite{noauthor_nlnet_nodate}, and Knot 5.7.4\cite{noauthor_knot_nodate}, each with a 100MB cache.
}

\hideforifip{ \rca{Shorthen - just refer to the mitigation section}
To address the vulnerabilities identified through the LAPRAD methodology, we propose the following mitigations:
\begin{itemize}
    \item Limit Oversized Signatures: RRSIG signature sizes should be capped based on the maximum size of supported algorithms, such as RSA-4096 (744 bytes). Exceptions can be made for domains requiring larger signatures. 
    \item Reduce DNSKEY Limits: Lower the max-records-per-type setting for DNSKEYs from the default (100) to a smaller value (e.g., 10) to reduce the number of RRSIGs per RRSet.
    \item ANY Query Restrictions: Implement separate limits for ANY-type queries to prevent large cumulative responses.

\end{itemize}

To develop effective mitigation techniques, we analyzed DNSSEC data from the Cloudflare top 100,000 domains\cite{cloudflare_top100k}. Among these, 9,336 were DNSSEC-enabled, and we examined the distribution of DNSKEY numbers and DNSSEC response sizes. These measurements provided valuable insights for shaping mitigation strategies. Using LLM assistance, we evaluated potential mitigations and confirmed their alignment with the techniques we manually proposed. The combined insights from these efforts aim to eliminate vulnerabilities while preserving DNS functionality.
}

\showforifip{

To develop effective mitigations for the newly discovered attacks, we conducted measurements on DNSSEC-enabled domains within the Cloudflare Top 100,000 list \cite{cloudflare_top100k}. Based on these observations, we identified practical limits and proposed mitigation preventing abuse of the DNSSEC mechanism.
}

\hideforifip{

To mitigate these attacks we propose limiting RRSIG sizes based on supported algorithms (e.g., RSA-4096, 744 bytes), reducing the max-records-per-type setting for DNSKEYs (from 100 to 20) to minimize excessive signatures, and setting specific limits on ANY queries to prevent large cumulative responses.
To set these limits, We examined the distribution of DNSKEY numbers and DNSSEC response sizes in DNSSEC-enabled domains within the Cloudflare top 100,000 domains\cite{cloudflare_top100k}.

}

\hideforifip{Using LLM assistance, we evaluated potential mitigations and confirmed their alignment with manually proposed mitigations.}

\hideforifip{
Section 2 provides background information on DNSSEC and the role of LLMs in this research. Section 3 outlines the threat model. In Section 4, we introduce the LAPRAD methodology, for discovering both known and zero-day attacks. Section 5 details the experiments conducted on the newly discovered  attacks. Section 6 reviews related work, Section 7 discusses findings and future work, and Section 8 concludes the paper.
}

\showforifip{
Section 2 provides background information on DNSSEC and the role of LLMs in this research, and related work. Section 3 outlines the threat model. In Section 4, we introduce the LAPRAD methodology, for discovering both known and zero-day attacks. Section 5 details the experiments conducted on the newly discovered  attacks. Section 6 discusses findings and future work, and Section 7 concludes the paper.
}


\section{Background and Related Work}

\hideforifip{\subsection{Large Language Models}
Large language models (LLMs) are advanced AI systems designed to excel in understanding and generating human-like text. They analyze and predict patterns in text using sophisticated deep learning techniques,  enabling them to capture context, meaning, and coherence. These models are trained on massive datasets and billions of parameters, enabling them to generate highly coherent and context-aware responses\cite{brown_language_2020}. Thanks to their ability to process and predict text, LLMs have proven successful in applications such as natural language understanding, conversational AI, creative writing, and code generation.

Prompt engineering—the practice of designing and structuring inputs to guide large language models (LLMs)—is used to enhance the performance and quality of their responses. In this paper, we employ several such techniques, including in-context learning\cite{dong2022survey}, where examples of a concept, such as network protocol RFCs, are provided within the input prompt; one-shot learning, where a single example guides the model’s understanding of a task \cite{brown_language_2020}; instructing the model to act in a specific role (e.g., 'you are an expert in the DNS protocol') \cite{kong2023better}; and Chain of Thought prompting \cite{wei_chain--thought_2023}, which explicitly guides the model to articulate logical steps when generating responses.


\subsection{DNS and DNSSEC}
DNS resolutions are initiated when client DNS requests are sent to \textit{recursive resolver servers}. These resolvers handle the name resolution process by querying from the root domain and traversing down the name hierarchy until they reach the \textit{authoritative nameserver}(auth) for the requested domain. Once the correct IP address for the queried name is obtained, this record is stored in the resolver’s cache and returned to the corresponding client. DNS responses include the answers in the form of \textit{RRSets} (Resource Record Sets), which logically separate different types of DNS data such as A (address), MX (email), NS (nameserver), and others. 

\textit{DNSSEC} protects the DNS system against unauthenticated answers and addressing vulnerabilities such as cache poisoning, where attackers inject malicious data into a resolver’s cache to redirect users to fraudulent websites. Each RRSet is signed by the domain’s \textit{DNSKEY}, ensuring the authenticity of responses, and the signature for each RRSet is stored in a corresponding \textit{RRSIG} record. The \textit{DNSKEY} is authenticated using the \textit{DS} (Delegation Signer) RRSet, which contains a hash of the \textit{DNSKEY} and is stored in the parent zone. The \textit{DS} RRSet is signed by the parent domain's key, creating a hierarchical chain of trust that extends to the root domain. Recursive resolvers, pre-configured with the root domain’s cryptographic public key (the trust anchor), use this chain of trust to verify the authenticity of DNS responses at every level.

DNSSEC-enabled domains use two types of keys: a Key Signing Key (\textit{KSK}) and a Zone Signing Key (\textit{ZSK}). The \textit{KSK} signs the DNSKEY RRSet, which contains the public keys used by resolvers, while the \textit{ZSK} signs other records in the zone. For example, when resolving an A-type RRSet, the \textit{KSK} is authenticated via its \textit{DS} record from the parent zone,\footnote{This process assumes the parent zone and higher levels in the hierarchy, up to the root domain, have already been authenticated.} and the \textit{ZSK} is used to authenticate the requested record. While most domains operate with these two keys, mechanisms such as key rollover and multi-signer configurations may allow a zone to use an unlimited number of keys under certain conditions.

}

\hideforifip{

\textit{\textbf{Large Language Models}} LLMs are AI systems that generate human-like text by analyzing patterns with deep learning. Trained on massive datasets, they capture context and meaning to produce coherent responses \cite{brown_language_2020}. \hideforifip{Their ability to process text makes them useful for tasks such as language understanding, conversational AI, writing, and code generation.} 

Prompt engineering structures inputs to improve LLM responses. For our prompts, we applied techniques such as in-context learning \cite{dong2022survey}, where prompts include examples (e.g., network protocol RFCs); one-shot learning, where a single example guides the model \cite{brown_language_2020}; role-based prompting (e.g., "You are a DNS expert") \cite{kong2023better}; Chain of Thought prompting, which helps the model outline logical steps \cite{wei_chain--thought_2023}; ReAct, which allows LLM agents to interact with the external environment, retrieve outputs, and adjust their actions accordingly \cite{yao2022react}; and the Multi-LLM agents approach, which involves multiple distinct LLMs collaborating to complete a task.


\textit{\textbf{DNS and DNSSEC}} DNS resolution begins when a client sends a request to a \textit{recursive resolver} to obtain the IP address of a domain. The resolver then queries the DNS hierarchy—from the root domain down to the \textit{authoritative nameserver} of the requested domain. Once the resolver retrieves the correct IP address, it stores the result in its cache and returns the IP address to the client. DNS responses include \textit{RRSets} (Resource Record Sets), which group related DNS record types, such as A (address) and NS (nameserver).

DNSSEC protects DNS from unauthenticated answers and cache poisoning attacks. Each \textit{RRSet} is signed by the domain’s \textit{DNSKEY}, ensuring response authenticity, with signatures stored in \textit{RRSIG} records. DNSSEC-enabled domains use two key types: the Key Signing Key (\textit{KSK}) and the Zone Signing Key (\textit{ZSK}). The \textit{KSK} is authenticated via the \textit{DS} (Delegation Signer) \textit{RRSet}, which contains a hash of the \textit{KSK}, is stored in the parent zone, and is signed by the parent domain’s key, forming a hierarchical chain of trust up to the root. Recursive resolvers, pre-configured with the root domain’s public key (trust anchor), use this chain to verify DNS responses. When resolving an RRSet in a zone, the \textit{KSK} verifies the DNSKEY RRSet, which contains the \textit{ZSK}, and the \textit{ZSK} verifies all the other RRSets (e.g., A, NS, MX).

\hideforifip{\footnote{Assuming the parent zone and higher levels up to the root domain are already authenticated.}}
}

\showforifip{

\textit{\textbf{LLMs}} are AI systems that generate human-like text by analyzing patterns with deep learning. Trained on massive datasets, they capture context and meaning to produce coherent responses \cite{brown_language_2020}. \hideforifip{Their ability to process text makes them useful for tasks such as language understanding, conversational AI, writing, and code generation.} 

Prompt engineering structures inputs to improve LLM responses. For our prompts, we applied techniques such as in-context learning \cite{dong2022survey}, where prompts include examples (e.g., network protocol RFCs); one-shot learning, where a single example guides the model \cite{brown_language_2020}; role-based prompting (e.g., "You are a DNS expert") \cite{kong2023better}; Chain of Thought prompting, which helps the model outline logical steps \cite{wei_chain--thought_2023}; ReAct, which allows LLM agents to interact with the external environment, retrieve outputs, and adjust their actions accordingly \cite{yao2022react}; and the Multi-LLM agents approach, which involves multiple distinct LLMs collaborating to complete a task.


\ignore{
\textit{\textbf{DNS and DNSSEC}}
\rca{I want to remove this part. Although it is good to prepare the reader for DNSSEC but taking a lot of space. - Maybe a very short referal to relevant RFCs without actually citing the RFCs because they are going to take space in the references part as well. }
}

\ignore{
\textcolor{red}{
DNS resolution begins when a client sends a request to a \textit{recursive resolver} to obtain the IP address of a domain. The resolver then queries the DNS hierarchy—from the root domain down to the \textit{authoritative nameserver} of the requested domain. Once the resolver retrieves the correct IP address, it stores the result in its cache and returns the IP address to the client. DNS responses include \textit{RRSets} (Resource Record Sets), which group related DNS record types, such as A (address) and NS (nameserver).}
}

\textit{\textbf{DNSSEC}} esures the autheticity of DNS responses, protecting from unauthorized responses, e.g., cache poisoning attacks. Each \textit{RRSet} is signed by the domain’s \textit{DNSKEY} (which is protected by a chain of trust-keys starting at the root NS), ensuring response authenticity, with signatures stored in \textit{RRSIG} records.
DNSSEC is inherently complex, as it incorporates redundancy, high availability, and strong security mechanisms.

\ignore{
\textcolor{red}{
DNSSEC-enabled domains use two key types: the Key Signing Key (\textit{KSK}) and the Zone Signing Key (\textit{ZSK}). The \textit{KSK} is authenticated via the \textit{DS} (Delegation Signer) \textit{RRSet}, which contains a hash of the \textit{KSK}, is stored in the parent zone, and is signed by the parent domain’s key, forming a hierarchical chain of trust up to the root. Recursive resolvers, pre-configured with the root domain’s public key (trust anchor), use this chain to verify DNS responses. When resolving an RRSet in a zone, the \textit{KSK} verifies the DNSKEY RRSet, which contains the \textit{ZSK}, and the \textit{ZSK} verifies all the other RRSets (e.g., A, NS, MX).}
}

\textbf{\textit{DNS DDoS Attacks}} \hideforifip{NXNSAttack amplifies traffic by manipulating referral responses between resolvers and authoritative servers \cite{afek2020nxnsattack}.
TsuNAME leverages cyclic dependencies in DNS records to cause recursive resolvers to loop, overloading servers \cite{moura_tsuname_2021}.}
NRDelegation exploits non-responsive authoritative responses to trigger excessive retries, depleting server CPU resources \cite{afek_nrdelegationattack_nodate}.
KeyTrap targets DNSSEC by sending malicious packets that force intensive cryptographic computations, resulting in CPU exhaustion \cite{heftrig_harder_2024}. 
CacheFlush\cite{afek2024flushing} attack exploited large resource record sets, packing up to 2,000 NS records into a single response, 65kB, per malicious query.
An attacker sending 1K qps (queries per second) easily flushes a 100MB resolver cache, reducing throughput by 80\%.
Different recent patches mitigate these including the latest, by capping RRSets at 100 records, rendering the CacheFlush attack infeasible with standard DNS RRSets.
\ignore{
\\\rca{below I suggest removing some cited work because they are redundant}
}

\textbf{\textit{DNS Protocol Modeling and Attack Discovery}} GRoot \cite{kakarla_groot_2020} introduced a formal model for DNS resolution to detect configuration errors such as rewrite loops, blackholing, and missing glue records.
\ignore{
\textcolor{red}{ SCALE \cite{kakarla_scale_2022} built upon this model, leveraging symbolic execution based on RFCs to generate test cases for differential testing across DNS implementations.}
}
Liu et al. \cite{liu_formal_2023} improved DNS modeling by formalizing resolver logic, including query state, and caching. However, these models provide a limited view of the actual DNS protocol.

\textbf{\textit{Applications of LLMs in Network Security}} LLMs have been applied to identify security issues in both general applications and network protocols. For example, they have been used to enhance fuzzing for IoT devices\cite{ma2024one} and to identify inconsistencies in 4G/5G documentation\cite{rahman2024cellularlint}. 
\ignore{
\textcolor{red}{remove---automate formal verification of 5G protocols \cite{wray20245g},} 
}
\ignore{
\textcolor{red}
{remove -- and detect vulnerabilities from GitHub issues \cite{cipollone2025automating}}.
}
Eywa \cite{kakarla_oracle-based_2023} used LLMs to generate partial DNS resolver functions for testing and finding vulnerabilities. In contrast, our approach uses protocol knowledge directly, making vulnerability identification easier without relying on implementation. Google used LLM agents to identify a previously unknown stack buffer underflow in SQLite by leveraging a relevant code change—and guiding the system to search for similar issues \cite{googleprojectzero2024}. This approach relates to our work as it trains LLMs on past examples to find similar vulnerabilities though it focuses on code-level issues.

}

\section{Threat Model}
\hideforifip{In our threat model, the attacker controls at least one client to issue malicious DNSSEC queries and an authoritative DNS server to host malicious DNSSEC domains, that respond with crafted replies containing large amounts of bogus data.

An attacker needs to purchase only a single domain, typically costing around \$1. Under this domain, the attacker can delegate multiple subdomains (e.g., sub1.attacker.com, sub2.attacker.com, etc.) at no additional cost, as they are all created under the same parent domain. In our case, a single malicious domain can inject approximately 200KB of junk data into the cache; thus, to completely flush a 100MB cache, an attacker needs at least 500 subdomains, while a 2GB cache may require up to 10,000 subdomains. Using 20,000 subdomains, however, is more efficient for flushing the cache with different attack domains, as this forces the resolver to fetch new domains from the attacker’s authoritative server. Additionally, bogus DNSSEC RRSIGs can be generated on the fly and included in the DNSSEC response, which eliminates the need for additional memory space for each RRset. As a result, increasing the number of subdomains does not significantly increase the attacker’s costs or memory usage. This threat model is common and has been employed in previous works such as NS CacheFlush\cite{afek2024flushing} and KeyTrap\cite{heftrig_harder_2024}.

The attacker can operate a malicious authoritative server on a limited budget using two main methods. First, the attacker can manage the server independently using cloud virtual machines, where the primary cost is outgoing traffic, estimated at approximately \$0.60 per minute during a 2GB cache-flushing attack. Second, the attacker can utilize managed DNS services from registrars such as Cloudns or GoDaddy, which often provide premium plans allowing unlimited subdomains and requests for about \$15 per month\cite{afek2024flushing} .}

\showforifip{
In our threat model, the attacker controls a client to send malicious DNSSEC queries and an authoritative server to host domains that return crafted replies with excessive bogus data. An attacker can purchase a single domain for about \$1 and create unlimited subdomains (e.g., sub1.attacker.com, sub2.attacker.com) at no extra cost. A single malicious domain can inject around 200KB of junk data into the cache, meaning 500 subdomains could flush a 100MB cache, while a 2GB cache may require 10,000. Using 20,000 subdomains is even more effective, as it forces the resolver to fetch new domains from the attacker's authoritative server. Bogus DNSSEC \textit{RRSIGs} can be generated on the fly without cryptographic operations, requiring no extra memory per \textit{RRSet}, so adding subdomains has minimal impact on cost or cpu/memory usage. This attack model is well-known and used in \textit{CacheFlush} \cite{afek2024flushing} and \textit{KeyTrap} \cite{heftrig_harder_2024}.

An attacker can run an authoritative server on a budget in two ways: hosting it on the cloud, with outgoing traffic costing around \$0.60 per minute for a 2GB cache-flushing attack, or using managed DNS services like GoDaddy’s premium plan, which offers unlimited subdomains and requests for \$15 per month \cite{afek2024flushing}.

}
\label{sec:threat_model}
\section{LAPRAD methodology for DNS}
\hideforifip{In this chapter, we describe the LAPRAD methodology for discovering DNS and DNSSEC vulnerabilities. We utilized GPT-o1\footnote{GPT-o1 was used for the primary attack discovery workflow, while GPT-4o was reserved for handling basic queries and secondary tasks.} and GPT-v4\cite{noauthor_gpt-4_nodate} as our LLM models for attack discovery, selected for their extensive training sets and advanced generative capabilities. Using this methodology, we discovered new DNSSEC CacheFlushing vulnerabilities and demonstrated its effectiveness in uncovering protocol-level flaws. Additionally, this technique was applied to re-discover two recent DNSSEC and DNS attacks, namely KeyTrap\cite{heftrig_harder_2024} and CacheFlush\cite{afek2024flushing}, which were not included in GPT's training data. These vulnerabilities were successfully identified with minimal or no hints provided to GPT. While we outline the key steps for the investigated attacks in this chapter, the full investigation steps for all attacks, along with the other conversations used in this paper, are available online \cite{all_gpt_conversations}. Our attack discovery methodology can also be applied to other protocols or domains, provided the LLM has been trained with the specifications of the targeted protocol or application."
}
\hideforifip{
In this section, we present the LAPRAD methodology for discovering DNS and DNSSEC vulnerabilities. We used GPT-o1 for attack discovery workflows and GPT-v4o for basic queries and secondary tasks \cite{noauthor_gpt-4_nodate}. This methodology uncovered new DNSSEC CacheFlushing vulnerabilities and effectively identified protocol-level flaws. It also re-discovered recent DNSSEC and DNS attacks, KeyTrap \cite{heftrig_harder_2024} and CacheFlush \cite{afek2024flushing}, which were not part of GPT’s training data. These vulnerabilities were found with little to no prior hints.  Due to space limitations, we outline only the key steps of the newly discovered SigCacheFlush attacks in this section. The full LLM conversations for all discovered vulnerabilities are available online \cite{all_gpt_conversations}.}

\showforifip{
Here we present the LAPRAD methodology for discovering DNS and DNSSEC DDoS vulnerabilities.
We used GPT-o1 for attack discovery workflows and GPT-v4o for basic queries and secondary tasks \cite{noauthor_gpt-4_nodate}.
We uncover new DNSSEC CacheFlushing vulnerabilities, and 
re-discovered recently published DNSSEC and DNS attacks — KeyTrap~\cite{heftrig_harder_2024} and CacheFlush~\cite{afek2024flushing} — which were not part of GPT’s training data.
Both papers were published after the October 2023 training cutoff date of the models used.
Due to space limitations, we outline only the key steps of the newly discovered SigCacheFlush attacks and briefly cover the Keytrap discovery.
The full LLM conversations for all discovered vulnerabilities are available online \cite{all_gpt_conversations}.}

\begin{figure*}[ht]
    \centering
    \includegraphics[width=\textwidth]{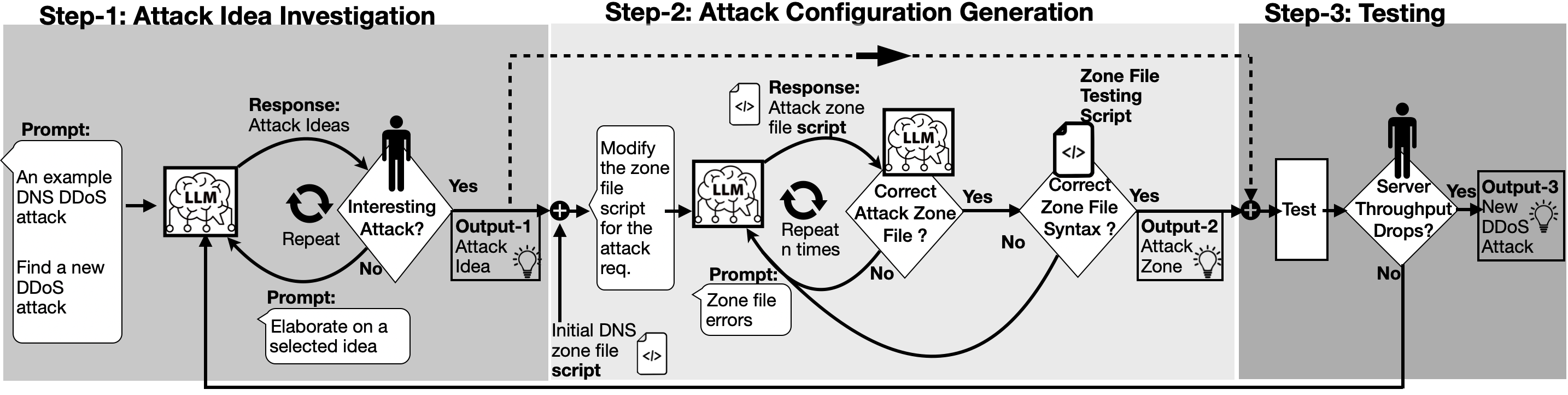} 
    \caption{Attack Investigation Workflow. The human figure indicates where human assistance is required to complete the task. The dashed line connects Output-1 (the attack idea) from \textit{\textbf{Step-1}} to \textbf{\textit{Step-3}}, illustrating that it will be tested using the attack zone configuration (Output-2).}
    \label{fig:investigation_workflow}
\end{figure*}

\subsection{LLM-Assisted Protocol Attack Discovery (LAPRAD)}

LAPRAD is a conversational, multi-step prompting-based strategy for attack investigation, designed to systematically leverage LLM assistance to discover vulnerabilities, as illustrated in Figure-\ref{fig:investigation_workflow}.
\hideforifip{

\begin{itemize}
\item \textbf{\textit{Step-1: Attack Idea Investigation }} 
            We instruct the LLM to act as a DNS security expert, providing a detailed example DNS-based DDoS attack description to guide its understanding. The LLM is then prompted to generate alternative DNS-based DDoS attacks targeting CPU, memory, or network bandwidth. An example initial prompt in this \textbf{\textit{Step-1}} is shown in {Prompt-1}. The LLM produces a list of attack ideas for a human researcher to review:
         \begin{itemize}
            \item  The researcher  manually selects one of the proposed attack ideas generated by the LLM.
            \item The researcher instructs the LLM to elaborate on the selected attack idea, refining it to make the attack more advanced. The goal in this step is to uncover protocol features that an attacker could exploit to enhance the effectiveness of the attack.
            \item This feedback loop is repeated for several rounds until the attack idea is sufficiently interesting for testing.
    \end{itemize}
    \item \textbf{\textit{Step-2: Attack Configuration Generation}} The selected attack idea, shown as \textbf{Output-1} in Figure-\ref{fig:investigation_workflow}, is combined with a benign zone generation script to create an attack zone generation script tailored to the attack requirements.
    \item \textbf{\textit{Step-3: Testing}}  Perform a test on the LLM-proposed attack using the dns zone file generated in \textbf{\textit{Step-2}}. If the attack is invalid or the researcher opts to try a different idea, the process returns to \textbf{\textit{Step-1}} to select another attack.
\end{itemize}
}

\showforifip{

\textbf{\textit{Step-1: Attack Idea Investigation }} 
We instruct the LLM to act as a DNS security expert, providing a detailed example DNS-based DDoS attack description to guide its understanding. The LLM is then prompted to generate alternative DNS-based DDoS attacks targeting CPU, memory, or network bandwidth. An example initial prompt in this \textbf{\textit{Step-1}} is shown in {Prompt-1}. The LLM produces a list of attack ideas for a human researcher to review:\\
- The researcher  manually selects one of the proposed attack ideas generated by the LLM.\\
- The researcher instructs the LLM to elaborate on the selected attack idea, refining it to make the attack more advanced. The goal in this step is to uncover protocol features that an attacker could exploit to enhance the effectiveness of the attack.\\
- This feedback loop is repeated for several rounds until the attack idea is sufficiently interesting to the researcher for testing.

\textbf{\textit{Step-2: Attack Configuration Generation}} The selected attack idea, shown as \textbf{Output-1} in Figure-\ref{fig:investigation_workflow}, is combined with a benign zone generation script to automatically create an attack zone generation script tailored to the attack requirements.\\
\textbf{\textit{Step-3: Testing}}  Perform a test on the LLM-proposed attack using the DNS zone file generated in \textbf{\textit{Step-2}}(\textbf{Output-2}). If the attack is invalid or the researcher opts to try a different idea, the process returns to \textbf{\textit{Step-1}} to select another attack.
}

\gptpromtbox{attack-prompt}{
\textbf{You are a DNS security expert.} Here is an example of a DNS attack that affects the cache of the resolver and reduces its performance:

\textbf{NS Cache Flush Attack}: An attacker owns an authoritative server and controls the example.com domain.
The attacker issues A-type queries for different unique names under this domain, such as test1.example.com, ..., test10000.example.com. The attacker’s authoritative server responds to these queries with approximately 1500 NS records, making the DNS reply reach the maximum allowed size of 65 KB. Each of these records is cached by the resolver, leading to the eviction of benign cached records. This forces the resolver to re-query these domains, significantly reducing its performance.

\textbf{Task}: Come up with another cache-flushing attack based on DNSSEC. Utilize your DNSSEC knowledge and explore potential mechanisms that could be used to target a resolver for cache flushing purposes.
}
\subsection{Discovering New DNS Vulnerabilities\ignore{ with LAPRAD}}
Here we use LAPRAD to discover new DNSSEC-based cache-flushing vulnerabilities and analyze their implications in three steps, see Figure-\ref{fig:investigation_workflow}. 
\textbf{\textit{Step-1}} described below, \textbf{\textit{Step-2}} in Section~\ref{subsubsec:gen-attack-configs}, and \textbf{\textit{Step-3}} in Section~\ref{sec:experiments}.
\ignore{
In this section, we detail the discovery of new DNSSEC-based cache-flushing vulnerabilities and analyze their implications, following the three steps outlined in Figure-\ref{fig:investigation_workflow}. \textbf{\textit{Step-1}} is described below, \textbf{\textit{Step-2}} is covered in Section~\ref{subsubsec:gen-attack-configs}, and \textbf{\textit{Step-3}} is discussed in Section~\ref{sec:experiments}.
}

\subsubsection{\textbf{Step-1: Attack Idea Investigation}}
\label{subsubsec:attack-idea-investigation}
We used the recently discovered \textit{CacheFlushing attack} \cite{afek2024flushing} as an illustrative example in Prompt-\ref{attack-prompt} and tasked the LLM with identifying novel methods to flush a DNS resolver's cache through DNSSEC features, ultimately uncovering the \textit{SigCacheFlush} vulnerabilities.

\ignore{This attack generates 65KB DNS responses by leveraging approximately 1,500 NS referral records.}

GPT responded to Prompt-\ref{attack-prompt} by suggesting an attack that combines multiple DNSKEYs and signatures, multiple DS records, and NSEC3 records. We focused on the aspect involving large DNSKEY records and multiple signatures, as shown in (GPT-Response-\ref{large_sig_dnssec_cacheflush}).\showforifip{ Then we used this answer to follow up with GPT using Prompt-\ref{question_max_rsa}.}

\gptanswerbox{large_sig_dnssec_cacheflush}{
    \hl{The attacker's domain is signed with multiple large DNSKEY records}. For instance, it can include: Multiple RSA-4096 keys, ..., or A mix of RSA/ECDSA/EdDSA algorithms... \hl{Each DNSKEY RRset might be tens of kilobytes when combined (especially if multiple RRSIGs are present)}.
}
\hideforifip{We used this answer to follow up with GPT using Prompt-\ref{question_max_rsa}:}
\gptpromtbox{question_max_rsa}{
Focus on GPT-Response-\ref{large_sig_dnssec_cacheflush}. How many RSA 4096 keys can fit into a 65KB DNS response, including all the RRSIGs as well?
}

\gptanswerbox{number_of_keys_answer}{
Each DNSKEY record and its associated RRSIG consume: Per-key size=600 bytes (DNSKEY)+400 bytes (RRSIG)=1,000 bytes per key ...\hl{The total number of DNSKEY records is: Number of keys=Available space/Per-key size= 65035/1000=65}(max-rsa packet-\cite{all_gpt_conversations})
}

\hideforifip{
The number of keys suggested in GPT-Response-\ref{number_of_keys_answer} was not entirely accurate (max-rsa packet-\cite{all_gpt_conversations})\hideforifip{\footnote{GPT suggested two different approaches in different conversations to reach the maximum number of keys per packet: either around 100 KSKs and 1 ZSK, resulting in 1 signature per RRSet, or around 60 ZSKs and 60 signatures. While the exact numbers of keys suggested may be incorrect, both approaches are valid.}}. We manually confirmed that a DNS response with 57 RSA-4096 DNSKEYs reaches 65KB.
}

The CacheFlushing patches \cite{CVE-2024-1737} set a default limit of 100 records per RRSet(BIND’s maximum-records-per-type setting).
GPT-Response-\ref{number_of_keys_answer} enables an attacker to stay within the 100-record limit while generating large responses.
We named this variation, \textbf{\textit{Multiple RSA-4096 Signatures-based Cache Flushing}} (multi-rsa\cite{all_gpt_conversations}).

To evaluate GPT's reasoning ability regarding this patch \cite{CVE-2024-1737}, we introduced a stricter limit and aimed to reach the maximum DNS packet size of 65KB using only two DNSKEYs instead of 100 (Prompt-\ref{patch_question}):

\gptpromtbox{patch_question}{
BIND has set a maximum limit of 2 records for any RRSet type. We need to be more creative to generate large responses containing 65KB answers in a single response, rather than multiple responses. Focus on any unbounded field that can help achieve this goal.
}

In GPT-Response-\ref{large_dnskeys}, GPT notices that DNSKEY and RRSIG RDATA fields are unbounded and proposes using two large DNSKEYs as a workaround for the Patch challenge.

\gptanswerbox{large_dnskeys}{
   ...nothing in the DNS wire-format truly prevents using absurdly large keys (e.g., 32k- or 64k-bit RSA).
    RRSIG RDATA: With a gigantic RSA key, the signature can also be thousands of bytes.
    \hl{In total, these 4 records can easily approach or exceed 60 KB if you size them aggressively. The DNS packet can reach the 65,535-byte maximum ...}
  \ul{  Resolvers can impose strict upper bounds on DNSKEY or RRSIG sizes and either truncate or refuse them.
... Some validating resolvers already treat extremely large keys as misconfigurations and fail validation.}
}

GPT-Response-\ref{large_dnskeys} states DNSSEC protocol specifications do not strictly limit signature size, though extremely large signatures may be rejected by resolvers. Then, we asked Prompt-\ref{accept_large_signature}.

\gptpromtbox{accept_large_signature}{
\ignore{In practice, }Would resolvers accept very large DNSKEYs or signatures?
}

\gptanswerbox{not_practical_large}{
...Even if a resolver does not explicitly reject large keys, \hl{it might run out of memory or trigger other internal limits }(e.g., “message size,” “maximum memory usage,” or “maximum TTL for suspicious data”). \hideforifip{Some software also enforces a maximum RRset size, beyond which the record set is truncated or dropped.}
}

GPT-Response-\ref{not_practical_large} states that large signatures are often impractical due to query timeouts from excessive memory overhead. As validation fails in such cases, we asked GPT the question in Prompt-\ref{no_validation}. \showforifip{Finally, GPT proposed an approach that bypasses the resolver validation and allows the resolver to cache the bogus signature in GPT-Response~\ref{bait_n_switch}.}

\gptpromtbox{no_validation}{
If we want to have a resolver accept a very large RRSIG without actually validating it, how can we achieve this?
}
\hideforifip{
We asked the same question three times, and GPT gave different answers each time. First, it suggested disabling DNSSEC for the resolver or marking the domain as insecure, assuming control over the recursive resolver—an incorrect assumption for our threat model. Next, it recommended disabling DNSSEC on the authoritative side. On the third attempt, GPT proposed a more advanced scenario after we ruled out these options, as outlined in GPT-Response-\ref{bait_n_switch}.

}

\gptanswerbox{bait_n_switch}{
... \hl{At least one of your DNSKEYs uses a mainstream algorithm} (e.g., RSA/SHA-256) that the resolver recognizes.\\
That legitimate key signs your zone normally, producing standard-sized RRSIGs that fully validate.\\
\hl{Using that huge unknown-algo DNSKEY, generate an enormous signature (}the RRSIG).\\
...at least \hl{one RRSIG from a recognized algorithm, the zone passes DNSSEC validation overall}. \hl{The “unknown algorithm” RRSIG is effectively ignored for validation.} The resolver sees it, doesn’t know how to validate it, but does not treat the zone as bogus.\\
    Typically, the \hl{resolver caches all RRs returned from the authoritative server—even those it did not use in successful validation}
}
\hideforifip{ GPT-Response-\ref{bait_n_switch} suggests that a resolver can cache a large DNSSEC signature (RRSIG) if the response contains at least one valid RRSIG for the RRSet. We named this attack the \textbf{\textit{Bait \& Switch CacheFlushing Attack}} (bait-discovery-\cite{all_gpt_conversations}).
\begin{figure*}[ht]
    \centering
    \includegraphics[width=0.75\columnwidth]{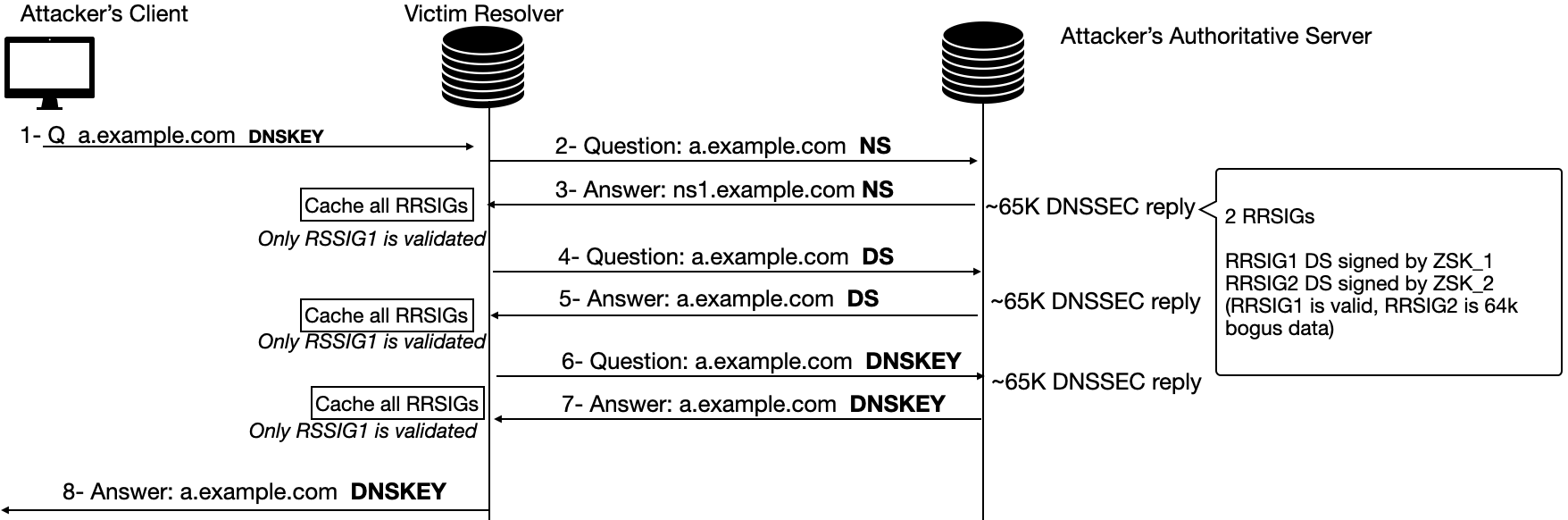} 
    \caption{Bait \& Switch Cache Flushing Attack}
    \label{fig:dnssec-attack}
\end{figure*}
}

\showforifip{
\begin{figure}[ht]
    \centering
    \includegraphics[width=\columnwidth]{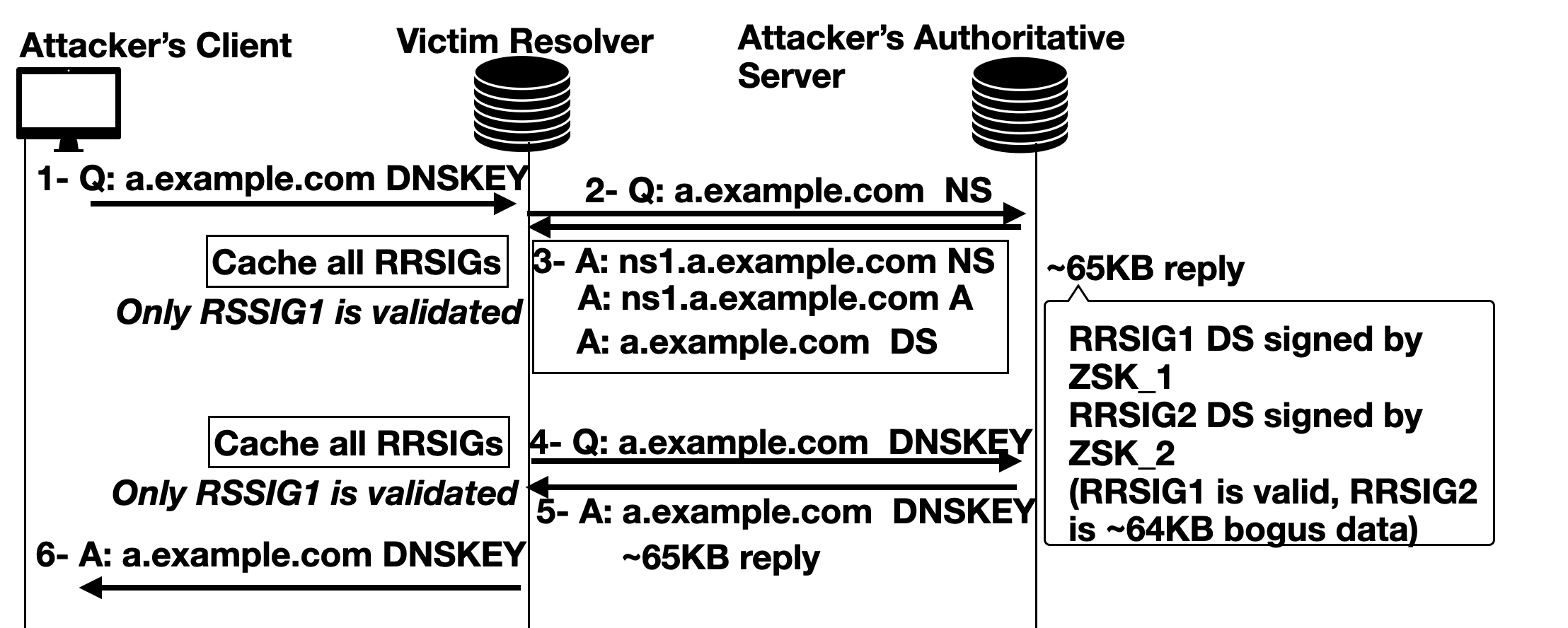} 
    \caption{Bait \& Switch Cache Flushing Attack}
    \label{fig:dnssec-attack}
\end{figure}

}

The idea described by GPT in GPT-Response-\ref{bait_n_switch} suggests that a very large DNSSEC signature (RRSIG) can be cached by the resolver, provided the response includes at least one valid RRSIG for the corresponding RRSet. We named this attack \textbf{\textit{Bait \& Switch CacheFlushing Attack}} (bait-discovery-\cite{all_gpt_conversations}). Figure~\ref{fig:dnssec-attack} illustrates the attack mechanism. In this example, the attacker responds with two RRSIG records: RRSIG1, a valid signature that the resolver successfully verifies, and RRSIG2, a bogus signature approximately 64KB in size. Although only RRSIG1 is validated, the resolver caches both signatures. Each query by the resolver—for the DS and DNSKEY RRSets—receives a 65KB response containing similar bogus data. As a result, a single malicious DNSKEY query causes roughly 130KB of bogus data to be inserted into the resolver’s cache. The \textit{SigCacheFlush} attack we discovered using LAPRAD bypasses the CacheFlush patches \cite{afek2024flushing}, which limit the number of NS records returned. Our attack leverages large DNSSEC signatures instead, making it unaffected by these mitigations.(\textit{Prompts for the ANY Type Cache Flushing Attack are omitted due to space, available at (any-discovery-\cite{all_gpt_conversations})}).

\hideforifip{We omitted the prompts related to  \textbf{\textit{ANY Type Cache Flushing Attack}} due to space constraints but full converstaion can be accessed online(any-discovery-\cite{all_gpt_conversations}).}

\refstepcounter{gptprompt} 
\label{skipped_box} 
\hideforifip{
In addition to these two attack ideas, GPT also suggested (GPT-Response-\ref{any_responses}) using ANY-type queries to maximize DNS response sizes for caching purposes\footnote{ Intermediate questions related to ANY-type queries (e.g., rompt-\ref{skipped_box}.) are omitted due to space constraints}. GPT reasoned that ANY-type responses could include multiple RRSet types, with each RRSet individually limited by the max-records-per-type cap. This approach enables bypassing the issued patch (default limit: max 100 RRSets) through a different method. We named this version of the attack the \textbf{\textit{ANY Type Cache Flushing Attack}}(\cite{any_investigation}).

\gptanswerbox{any_responses}{
\hl{BIND applies its maximum-records-per-RRset limit on a per-RRset basis,} including each RRSIG RRset individually. Thus, RRSIG(DS), RRSIG(DNSKEY), RRSIG(NS), etc. each get capped separately.\\
...\\
Even with BIND's per-RRset record limit (e.g., 100 records), an \hl{ANY query still aggregates multiple RRsets-each with up to 100 RRSIGs}. This allows an attacker to assemble a near-65kB response in one go.\\
}

}

\hideforifip{The exact number of user prompts(NP) and number of unique GPT-generated ideas per prompt ((NIP) in \textit{\textbf{Step-1}} in order to discover each attack idea is omitted due to space limitation, however is between $2$ and $7$, and $1$ to $8$ correspondingly.}

\hideforifip{The number of prompts to discover an attack (NPD) and unique GPT-generated ideas per prompt (NIP) in \textit{\textbf{Step-1}} are: \textbf{Multiple RSA} (4, 1), \textbf{Bait and Switch} (4, 1, 4, 4, 1, 1), \textbf{ANY Type attack} (4, 8, 1, 1), \textbf{Cache Flush-CNAME} (3, 4, 4, 4, 1), \textbf{KeyTrap} (5, 1, 1, 1, 8, 5, 1), with NPD ranging from $2$ to $7$.} 

\showforifip{
The number of prompts to discover an attack (NPD) and unique GPT-generated ideas per prompt (NIP) (formatted as \textbf{(}\# of ideas for 1st prompt, 2nd prompt,...\textbf{)}) in \textit{\textbf{Step-1}} are: \textbf{Multiple RSA} (4, 1), \textbf{Bait and Switch} (4, 1, 4, 4, 1, 1), \textbf{ANY Type attack} (4, 8, 1, 1), \textbf{Cache Flush-CNAME} (3, 4, 4, 4, 1), \textbf{KeyTrap} (5, 1, 1, 1, 8, 5, 1), with NPD ranging from $2$ to $7$.

}

\hideforifip{
We shared a detailed breakdown of number of prompts and responses iterated in \textit{\textbf{Step-1}} that is required to discover each attack idea in Table-\ref{tab:attack_ideas_numbers}.
}

\hideforifip{
\begin{table}[htbp] 
    \begin{tabularx}{\columnwidth}{|l||p{1.2cm}|X|}
        \hline
        \textbf{Attack Idea} & \textbf{NPD} & \textbf{NIP} \\  \hline
        Multiple RSA          & 2  & (4, 1)                    \\  \hline
        Bait and Switch       & 6  & (4, 1, 4, 4, 1, 1)       \\  \hline
        ANY Type responses    & 4  & (4, 8, 1, 1)             \\  \hline
        KeyTrap               & 7  & (5, 1, 1, 1, 8, 5, 1)    \\  \hline
        Cache Flush - CNAME   & 5  & (3, 4, 4, 4, 1)          \\  \hline
    \end{tabularx}
\caption{Summary of Attack Ideas and Discovery Metrics: This table lists the user prompts required to discover each attack (\textbf{NPD}) and the unique GPT-generated ideas per prompt (\textbf{NIP}), formatted as (\# of ideas for 1st prompt, 2nd prompt,...)}
\label{tab:attack_ideas_numbers}
\end{table}
}

\subsubsection{\textbf{Step-2: Attack Configuration Generation}}
\label{subsubsec:gen-attack-configs}

\hideforifip{
After GPT identified a potential attack, we asked it to generate detailed attack steps and configurations, including the attacker’s zone file. While GPT successfully outlined the steps, the generated zone files were inconsistent and too abstract for direct implementation. For a human reader, these files provided high-level guidance but lacked the detail necessary to run on a real server.

 To address this issue, we reframed the task as a code modification problem, leveraging the LLM's strength in code generation. As shown in Figure-\ref{fig:investigation_workflow}, we provided GPT with a baseline script that generates a zone file containing one KSK, one ZSK, and a signed zone file (one-shot learning), along with the selected attack idea.\rca{modify here}We then iteratively prompted GPT to refine the script based on the attack description until we were satisfied with the generated zone files. This approach enabled us to automatically generate accurate attacker zone files tailored to the discovered attack.
}

\showforifip{
We used GPT to generate detailed attack steps and configurations, including the attacker’s zone file. Although GPT successfully outlined the steps, the generated zone files were inconsistent and too abstract for direct implementation. To address this issue, we implemented a ReACT-based script to automatically generate zone files without human intervention (Figure-\ref{fig:investigation_workflow}, Step-2).

(2.1) The attack idea's zone requirements from Figure-\ref{fig:investigation_workflow}, Step-1, are passed to the ReACT script, which then calls the \textit{zone generation script} to create a basic DNS zone (e.g., sub.example.com). The LLM then checks if the zone file meets attack requirements, modifying the script iteratively for up to n = 5 iterations until the requirements are satisfied.\\
(2.3) Once the attack zone is structured, the ReACT script runs named-checkconf for syntax and rule validation. If errors occur, they are reported to the LLM, which updates the \textit{zone generation script} to fix them.\\
(2.4) After validation, the generated zone file is uploaded to a virtual machine. The ReACT script then checks the BIND server startup logs for errors. If issues arise, the \textit{zone generation script} is sent back to the LLM with the BIND errors for further corrections. These steps are combined in Step-2, as the \textit{Zone File Testing Script}.\\
(2.5) The final version of the \textit{zone generation script} is run in a loop to generate thousands of zone files, each with a given domain name as a parameter, ensuring sufficient attack subdomain configurations based on the victim resolver's size.
}

\subsubsection{\textbf{Source of Information}}
\label{subsubsec:source-of-info}

\hideforifip{
We investigated the sources of information for the newly discovered vulnerabilities. First, we asked GPT how it formulated a specific idea, such as using a small valid signature alongside a large bogus one, requesting the exact RFC number and a verbatim quote. GPT provided reasoning, citing relevant RFCs. Although the responses were logically correct and the RFC numbers were accurate, the quotes were hallucinated. We then incorporated GPT’s answer into another prompt with the attack summary and full RFC text, asking it to find and quote related sentences\ignore{ as shown in Figure-\ref{fig:source_info_workflow}}. This time, GPT successfully identified  relevant sentences in the given RFC.
}
\showforifip{
We investigated the sources of information for the newly discovered vulnerabilities in three steps. First (\circled{1}), we asked GPT how it formulated a specific idea, such as using a small valid signature alongside a large bogus one, explicitly requesting the exact RFC number and a verbatim quote. GPT provided reasoning and accurate RFC numbers, but the quotes were hallucinated. Next (\circled{2}), we combined this initial response + (\texttt{Attack Summary}) + (\texttt{Full RFC Text}) $\rightarrow$ Instructed GPT to find relevant sentences. Finally (\circled{3}), GPT correctly identified the sentences within the given RFC.
}

\hideforifip{We also investigated the source of information for the newly discovered vulnerabilities, following the steps outlined in Figure-\ref{fig:source_info_workflow}. First, we asked GPT how it developed a specific idea, such as using one small but valid signature alongside one bogus but very large signature. We requested the exact RFC number and a word-for-word quote of the relevant sentence. GPT responded with reasoning and explanations, citing the relevant RFCs and specific sentences it referenced to develop the idea.

Although the RFC and section numbers provided by GPT were partially correct, the exact sentences were generated rather than directly quoted from the RFCs. This behavior is expected, given that LLMs are trained to predict the next words in a sequence rather than memorize their training data verbatim. Despite this limitation, the generated sentences were logically accurate and effectively guided us to the correct information about the vulnerability.

We incorporated this response in our Prompt-\ref{source-info-prompt}, marked in red, along with the attack summary and the full version of the referenced RFC. We then asked GPT to find and quote any sentences related to the information in its original response. In this step, GPT successfully identified two relevant sentences, which we had manually verified. At the end of this process, we obtained the following sentences as the GPT output in GPT-Response-\ref{rfc_output}. These sentences are exact excerpts from the RFCs.

\gptpromtbox{source-info-prompt}{
\textbf{Summary of the Attack:} ....Since the goal is to get a large RRSIG cached without failing validation, we must provide at least one valid signature that the resolver can confirm, allowing the RRset to be accepted and cached.

\textbf{ Quoted Sections and RFCs: } Exact Citation: RFC 4035, Section 5.3.3 (Authenticating DNS Responses)\\
"\ul{If the resolver obtains RRSIG RRs and DNSKEY RRs for an RRset, the resolver SHOULD attempt to verify the signatures until it finds at least one that validates the RRset successfully, or until all the available RRSIG RRs have been found to fail.}"\\
\textbf{Reasoning:} .....After at least one signature validates, the RRset is considered secure, and the validator does not need to invalidate or discard the other RRSIGs—no matter their size\\
\textbf{File:} RFC 4035\\
\textbf{Task}: This is the reasoning provided by GPT for an attack it identified. However, the RFC references and sentences are not entirely accurate. I would like you to find the exact sentences and references that GPT was attempting to mention in the uploaded RFCs.
}

\gptanswerbox{rfc_output}{
RFC 6840 - 5.4 Caution about Local Policy and Multiple RRSIGs: This document specifies that a resolver SHOULD accept any valid RRSIG    as sufficient, and only determine that an RRset is Bogus if all    RRSIGs fail validation.\\
RFC 4035 4.5 Response Caching:  A security-aware resolver SHOULD cache each response as a single    atomic entry containing the entire answer, including the named RRset    and any associated DNSSEC RRs.
}

}
\ignore{
\begin{figure}[ht]
    \centering
    \includegraphics[width=\columnwidth]{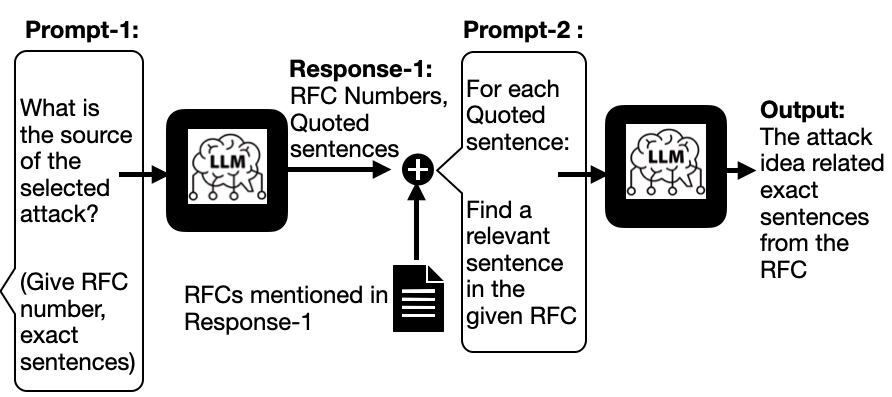} 
    \caption{Workflow for Tracing the Source of Proposed Attack Information}
    \label{fig:source_info_workflow}
\end{figure}
}

\subsubsection{\textbf{Proposed Mitigations}} \label{subsubsec:proposed-mitigation}

\hideforifip{We leveraged LLMs to identify potential mitigations, as they can synthesize protocol features and RFC details to generate both attack ideas and defenses. For each CacheFlushing attack type, we provided the attack description generated by LAPRAD as input to separate GPT conversations using Prompt-\ref{mitigation_prompt}. The mitigations proposed by GPT were compiled into a combined response, which is presented in GPT-Response-\ref{mitigation_answer} (see Appendix-\ref{appendix:A}). Below, we present the mitigation options we manually developed, which closely align with those suggested by GPT, along with additional ideas proposed by the models.}

\hideforifip{
We leveraged LLMs to identify potential mitigations, as they can synthesize protocol features and RFC details to generate both attack ideas and defenses. For each CacheFlushing attack type, we provided the attack description generated by LAPRAD as input to a separate GPT conversation and asked for mitigation strategies.\ignore{ Below, we present the mitigation options we manually developed, which closely align with those suggested by GPT(dnssec-mitigations-\cite{all_gpt_conversations}), along with additional ideas proposed by the models.}

Following are mitigation proposals for the newly discovered SigCacheFlush attacks. 
The following mitigations are being discussed in a responsible disclosure with several different vendors.\\

}

\textit{Bait \& Switch Cache Flushing Attack mitigation}: Prevent oversize responses by limiting RRSIG signature size, with the default cap set to the maximum signature size of supported algorithms, RSA-4096 (744 bytes).
\hideforifip{
This limit can be expanded for specific domains that require special algorithms.
}

\textit{Multiple RSA-4096 Signatures-based Cache Flushing Attack mitigation}: Lower the default max-records-per-type for DNSKEYs, and for RRSIGs per RRSet, from 100 to 20. Analyzing Cloudflare’s top 100,000 domains, we found 9,336 DNSSEC-enabled domains. As shown in Figure-\ref{fig:combined_measurements} b),
none of which require more than $16$ DNSKEYs.
\ignore{
98\% of which require at most 4 DNSKEYs (and 4 RRSIGs per RRSet), while 99.9\% need no more than 10 RRSIGs.
}
Setting a default limit to 20 would considerably reduce the effect of the attack. 

\textit{ANY Type Cache Flushing Attack mitigation}: ANY-type queries can trigger large cumulative responses because each RRSet type’s RRSIG is treated as a distinct RRSet, with limits applied separately. This allows numerous signatures (e.g., 100 for NS RRSIG and 100 for MX RRSIG). Setting separate limits for ANY-type queries can prevent this.

\showforifip{\begin{figure}[ht]
    \centering
    \includegraphics[width=\columnwidth]{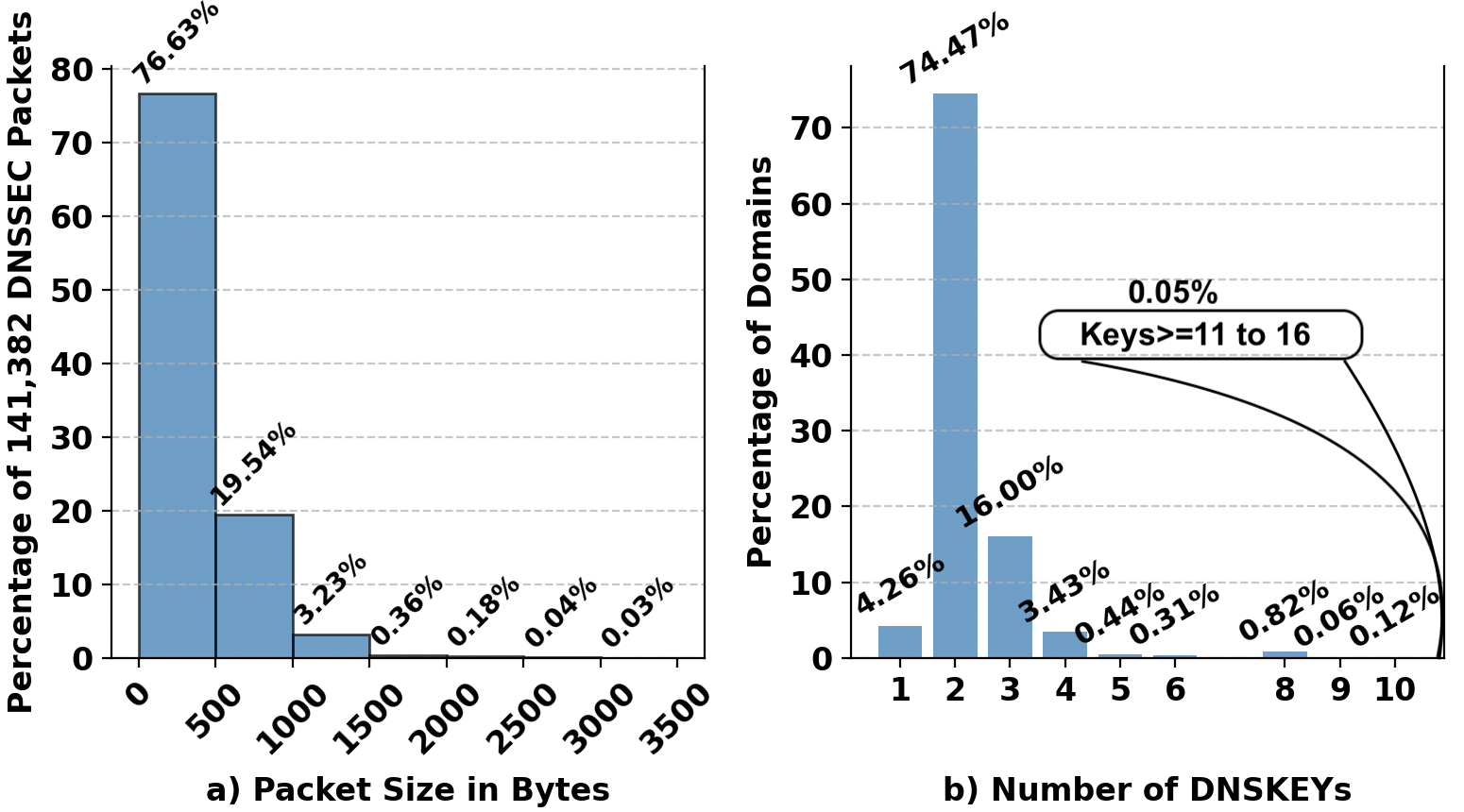} 
    \caption{For the top 9,336 DNSSEC-enabled domains and their TLDs, panel (a) 
    presents the DNSSEC response packet size distribution,
    while panel (b)
    shows the DNSKEY distribution (up to 16 keys per domain, with values from 11 to 16 keys omitted as they account for only 0.05\% of domains).
    }
    \label{fig:combined_measurements}
\end{figure}}
\hideforifip{
\subsection{Re-discovering Existing DNS Vulnerabilities\ignore{ with LAPRAD}}
LAPRAD assisted also in the re-discovery of Keytrap \cite{heftrig_harder_2024} and CacheFlush(2024) \cite{afek2024flushing} attacks.

We first confirmed GPT's knowledge cutoff as October 2023 by testing its awareness of publicly available information released afterward. After that, we checked whether it recognized the KeyTrap attack \cite{heftrig_harder_2024} (published February 2024) or the CacheFlush paper \cite{afek2024flushing} (released August 2024). In both cases, it showed no awareness, confirming its knowledge cutoff.

\textbf{\textit{KeyTrap attacks}} \cite{heftrig_harder_2024} exploit DNSSEC’s ‘Keytag collusion’ flaw. A recursive resolver validates an RRSIG by matching it with its corresponding key using the KeyTag value. However, since KeyTags are not necessarily unique, multiple keys can share the same value. DNSSEC requires validating all signatures in an RRSet until one succeeds, meaning all keys with the same KeyTag value must be tried. Attackers exploit this by sending \textit{n} invalid signatures with \textit{m} keys, forcing the validator to attempt all key-signature combinations before rejecting them, resulting in \textit{n×m} cryptographic operations.

LAPRAD has re-discovered the KeyTrap attack (keytrap-discovery-\cite{all_gpt_conversations}).
In \textbf{\textit{Step-1}}, we used a modified version of Prompt-\ref{attack-prompt}, replacing the example attack with the NRDelegation attack \cite{afek_nrdelegationattack_nodate}, which leverages non-responsive name-server referrals to increase the resolver computational complexity.
Following the initial prompt, GPT suggested several attack vectors. One of these involved using large DNSKEY sets to overwhelm the recursive resolver (highlighted in GPT-Response-\ref{keytrap_answer_1}).


\gptanswerbox{keytrap_answer_1}{
\hl{
... Large DNSKEY Sets}: Each DNSKEY must be verified and can consume significant CPU time, especially for \hl{large key sizes or expensive algorithms} (e.g., RSA....

}

We followed this recommendation (GPT-Response-\ref{keytrap_answer_1}) and explored it further in {Prompt-\ref{keytrap_prompt_2}}. In GPT-Response-\ref{keytrap_answer_2}, GPT suggested using multiple signatures and DNSKEYs in a single response to maximize the attack's impact.

\gptpromtbox{keytrap_prompt_2}{
Focus on large number of DNSKEY approach. Elaborate on this idea 
}
\gptanswerbox{keytrap_answer_2}{
... A resolver must verify each RRSIG with the corresponding public key (DNSKEY).
\hl{If a zone operator publishes, for example, dozens (or even hundreds) of DNSKEY records, the resolver must perform dozens (or hundreds) of signature verifications whenever it fetches the DNSKEY RRset.}
...\\
...In practice, you might see anywhere from 80 to 100+ pairs if you really push the 65 KB boundary with medium (2048-bit) RSA keys.
}
In Prompt-\ref{keytrap_prompt_3}, we examined the number of validations required for 100 signatures and 100 keys and explored ways to increase the validation count.

\gptpromtbox{keytrap_prompt_3}{

\textbf{User Prompt:} Let's say 100 DNSKEYs + RRSIGs. What is the number of validations the resolver need to perform for such DNSSEC reply?
\\
\textbf{GPT Response:} \hl{100 DNSKEYs = Up to 100 Validations}\\
\textbf{User Prompt:}\ul{ Is there way to increase the number validations for this DNSSEC reply?}
}
After this, GPT identified that KeyTag collisions could be exploited as an attack vector(GPT-Response-\ref{keytrap_answer_3}).

\gptanswerbox{keytrap_answer_3}{
"...How it Works: \hl{Give multiple DNSKEYs the same key tag}. Key tags are 16-bit values calculated from the public key data. While collisions are rare in normal practice,\hl{ an attacker can engineer collisions by carefully crafting keys.}
}

We prompted GPT to elaborate on this attack over two additional iterations, uncovering the core concept of the KeyTrap attack. This attack exploits multiple colliding keys with identical key tags and invalid signatures, forcing the resolver to attempt verification of all signatures against all keys.

\gptpromtbox{keytrap_prompt_4}{
GPT-Response-\ref{keytrap_answer_3}
Focus here. Elaborate on this idea. Lets think step by step
}

\gptanswerbox{keytrap_answer_5}{
\hl{\textbf{Repeated Signature Validation Attempts:}} ... scenario with 100 DNSKEYs and 100 corresponding RRSIGs, a resolver could be driven to perform up to 10,000 individual signature validation checks
\hl{ 100 DNSKEYs:Each DNSKEY has the same key tag }or otherwise appears to be a valid 
candidate (e.g., due to multiple DS records referencing all of them).\\
\hl{100 RRSIGs: Each RRSIG claims to sign the DNSKEY RRset} (or some relevant RRset), 
\hl{Trying Each DNSKEY Against Each RRSIG}:In a naive (or maliciously forced) validation process, the resolver sees RRSIG 1 and tries all 100 DNSKEYs before determining which one (if any) actually matches. Then it moves on to RRSIG 2, tries all 100 DNSKEYs again, and so on  through RRSIG 100.
Multiplying the Checks: \hl{100 RRSIGs x 100 DNSKEYs = 10,000 signature validations}....
}

\hideforifip{
\textbf{\textit{CacheFlushing attacks}}\cite{afek2024flushing}: Since our initial prompt for the KeyTrap attack instructed GPT to identify not only computational complexity attacks but also those involving excessive memory utilization, GPT recommended a cache flushing attack as follows (\cite{noauthor_cachefloodinginvestigation_nodate}):

\gptanswerbox{cache_flush_cname_answer_1}{
\hl{Signature Flood Attack}: This attack leverages the requirement for resolvers to store multiple valid signatures for a single DNS record, loading the resolver's cache and potentially its processing capabilities.\\
Detailed Breakdown:\\
 - Cache Flooding: The attacker generates\hl{ multiple valid DNSSEC signatures for the same records and has them cached by the resolver.}\\
- Memory Exhaustion: Storing multiple signatures requires more memory, which can degrade the resolver's performance and potentially lead to service denial if paired with other attacks.
}
Continuing with our attack discovery strategy, GPT recommended one of the CacheFlushing techniques described in \cite{afek2024flushing}, specifically leveraging the CNAME RRType.
}

\showforifip{
\textbf{\textit{CacheFlushing attacks:}} Since we prompted GPT to identify computational and memory-intensive attacks for Keytrap attack, it suggested a CacheFlushing technique using CNAME RRType \cite{afek2024flushing}. Due to space limitations, related prompts are omitted and can be found in \cite{all_gpt_conversations}.
}
}

\ignore{
\rca{I brought this part back.-- although very short version compared to the initial one. Only showing the last GPT response.}
}
\vspace{-0.5cm}
\ignore{
\textcolor{blue}
}
\subsection{Re-discovering Existing DNS Vulnerabilities\ignore{ with LAPRAD}}

\ignore{
}
\textbf{\textit{KeyTrap attack}} \cite{heftrig_harder_2024} exploits DNSSEC's KeyTag collusion flaw, where multiple keys match the same KeyTag. Since resolvers must try all matching keys until one validates the signature, attackers can force excessive computations by triggering these collisions.

\ignore{
}
LAPRAD re-discovered the KeyTrap attack (keytrap-discovery\cite{all_gpt_conversations}) in $7$ prompts.
We replace the example in Prompt~\ref{attack-prompt}, in Step-1, with the NRDelegation attack~\cite{afek_nrdelegationattack_nodate}, following the refinement process in Figure~\ref{fig:investigation_workflow}, Step-1. Due to space limitation, we only show the final GPT response below.

\gptanswerbox{keytrap_answer_5}{
\hl{\textbf{Repeated Signature Validation Attempts:}} ... scenario with 100 DNSKEYs and 100 corresponding RRSIGs, a resolver could be driven to perform up to 10,000 individual signature validation checks
\hl{ 100 DNSKEYs:Each DNSKEY has the same key tag }or otherwise appears to be a valid 
candidate (e.g., due to multiple DS records referencing all of them).\\
\hl{100 RRSIGs: Each RRSIG claims to sign the DNSKEY RRset} (or some relevant RRset), 
\hl{Trying Each DNSKEY Against Each RRSIG}:In a naive (or maliciously forced) validation process, the resolver sees RRSIG 1 and tries all 100 DNSKEYs before determining which one (if any) actually matches. Then it moves on to RRSIG 2, tries all 100 DNSKEYs again, and so on  through RRSIG 100.
Multiplying the Checks: \hl{100 RRSIGs x 100 DNSKEYs = 10,000 signature validations}....
}

\section{Experiments}
\label{sec:experiments}
\showforifip{\begin{figure*}[ht]
    \centering
    \includegraphics[width=\textwidth,height=0.25\textheight]{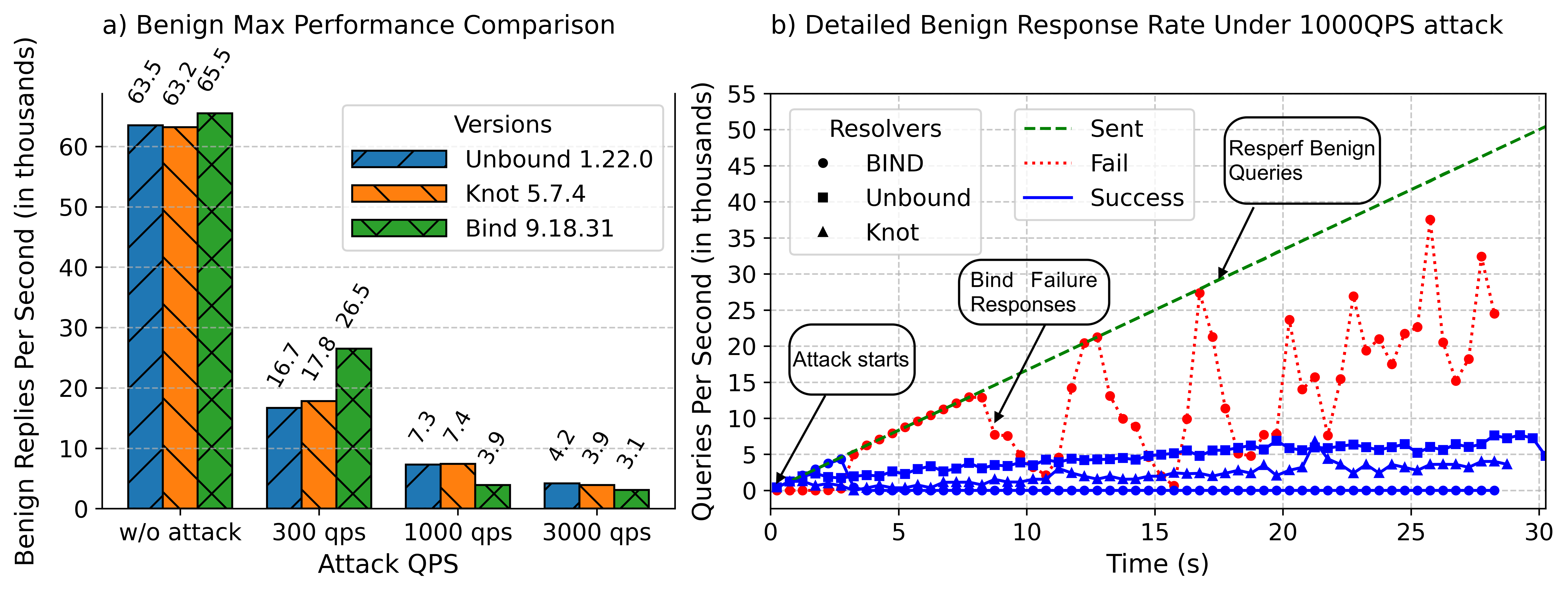} 
    \caption{a)Resolver Benign Max QPS Performance with Bait\&Switch Attack (excluding SERVERFAIL responses) under constant attack rates of 0, 300, 1,000, and 3,000 QPS. b) Detailed benign response rates of resolvers benign max QPS under a bait-and-switch attack at constant 1,000 QPS. The attack starts at 0s, marked by the black arrow.}
    \label{fig:max_qps_detail}
\end{figure*}}

\hideforifip{We conducted proof-of-concept tests on the three new attack ideas discovered in a Docker environment. These tests had two objectives: to verify whether the LAPRAD-proposed ideas could generate 65KB DNSSEC responses and to determine whether resolvers would accept them. The results showed that all three attack ideas generated near-65KB DNSSEC responses, which were accepted by the tested resolvers, confirming their feasibility. Since all three attacks aim to flush the resolver's cache using large responses, we focused our performance evaluation on the \textbf{\textit{Bait-and-Switch attack}}, which is the most intricate of the three.
}

\showforifip{
We tested the three newly discovered attacks in a Docker environment to verify that they generate 65KB DNSSEC responses accepted by BIND, Unbound, and Knot. All three produced responses close to 65KB, which were successfully cached by the tested resolvers, confirming their feasibility. \\\showforifip{We analyzed DNSSEC responses from 9,336 domains and found an average packet size of 449 bytes(Figure-\ref{fig:combined_measurements}, a). SigCacheFlush attacks amplify this by 144×.} Since all three attacks aim to flush the resolver's cache using large responses, we focused our performance evaluation on the \textbf{\textit{Bait-and-Switch attack}}, which is the most intricate one.
}

\textbf{\textit{Step-3: Testing (Figure-\ref{fig:investigation_workflow})}}
A human expert gets a; detailed attack idea, high-level testing guide from the LLM, and an attack zone file to complete the test.

The performance impact of the \textbf{\textit{Bait-and-Switch attack}} was evaluated in an Azure cloud environment. The setup included an attacker's authoritative server, a benign authoritative server, a recursive resolver, an attacker client, and a benign client. Each machine had an Intel Xeon E5-2673 v4 @ 2.30GHz (2 vCPUs), 8 GiB RAM, and Ubuntu 20.04\hideforifip{operating system (Figure-\ref{fig:test_environment})}.

\hideforifip{
\rca{this figure can be removed}
\begin{figure}[ht]
    \centering
    \includegraphics[width=\columnwidth]{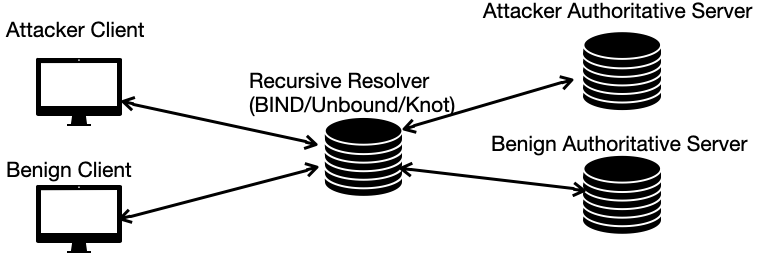} 
    \caption{Test Environment}
    \label{fig:test_environment}
\end{figure}
}

\hideforifip{
\subsection{Measurement of Packet Size Amplificatfion}
To evaluate the packet size amplification factor, similar to the measurements conducted in Section-\ref{subsubsec:proposed-mitigation}, we measured the average DNSSEC response size for the top 9,336 DNSSEC-enabled domains. Our analysis revealed an average packet size of 449 bytes. The histogram showing the distribution of packet sizes is presented in Figure-\ref{fig:dnssec_pkt_size_distro}. The discovered SigCacheFlushing attacks, which result in DNS packet sizes of up to 65 KB, demonstrate a 144X packet size amplification factor.}

To evaluate resolver performance under a \textbf{\textit{Bait-and-Switch attack}}, we use the Resperf \cite{resperf} tool, a widely accepted standard for measuring resolver throughput. In this test, we measured the resolver’s maximum throughput by using two query files: the attacker’s Resperf query file contained 10,000 unique subdomains under a single purchased domain (see Section~\ref{sec:threat_model} for cost calculation), while the benign query file contained 100,000 unique domain names. Both clients began with an empty cache: the benign client gradually increased its query rate until the resolver reached its capacity, while the attacker client maintained a steady query rate.

We tested the attack using three attacker query rates: 300 qps, 1,000 qps, and 3,000 qps. The tests were performed on the latest versions of three resolvers—BIND 9.18.31, Unbound 1.22.0, and Knot 5.7.4—all configured with a 100MB cache. Each test was repeated three times, and the average results are shown in Figure-\ref{fig:max_qps_detail}, a) . Resolver throughput is reported as the maximum number of benign queries per second.

During the BIND tests, we observed a transient peak of approximately 4,000 benign QPS for the 1,000 QPS attacker. However, once the attack took effect, all benign responses from the resolver resulted in 'SERVERFAIL' (Figure-\ref{fig:max_qps_detail}, b)).
In contrast, both Unbound and Knot experienced a significant drop in benign QPS under attack. Notably, Unbound never returned 'SERVERFAIL,' while Knot occasionally did, but at a rate of less than 1\%.

\hideforifip{
\begin{figure}[ht]
    \includegraphics[width=\columnwidth,height=0.20\textheight]{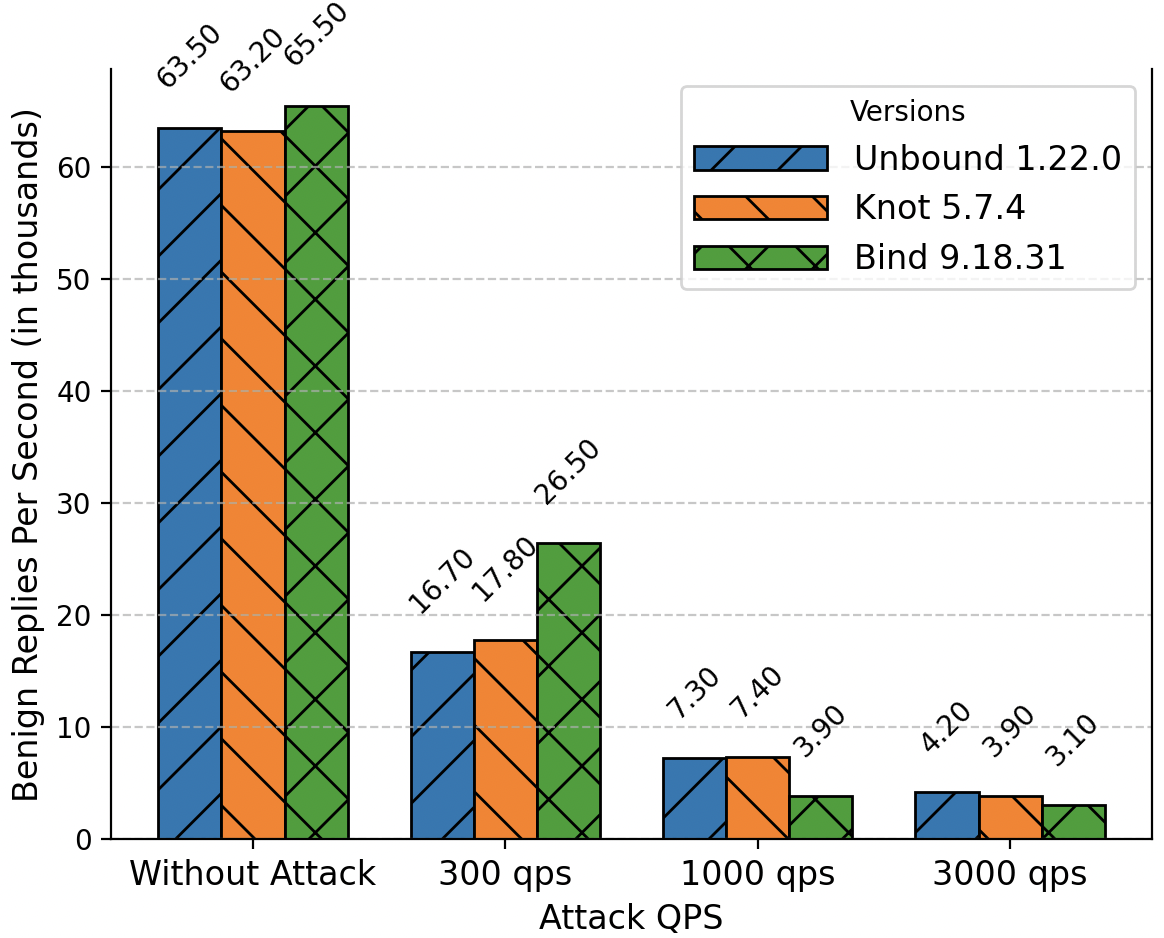} 
    \caption{Resolver Benign Max QPS Performance with Bait\&Switch Attack (excluding SERVERFAIL responses) under constant attack rates of 0, 300, 1,000, and 3,000 QPS.}
    \label{fig:performance_evaluation}
\end{figure}
}

\hideforifip{
\begin{figure}[ht]
    \centering
    \includegraphics[width=\columnwidth,height=0.25\textheight]{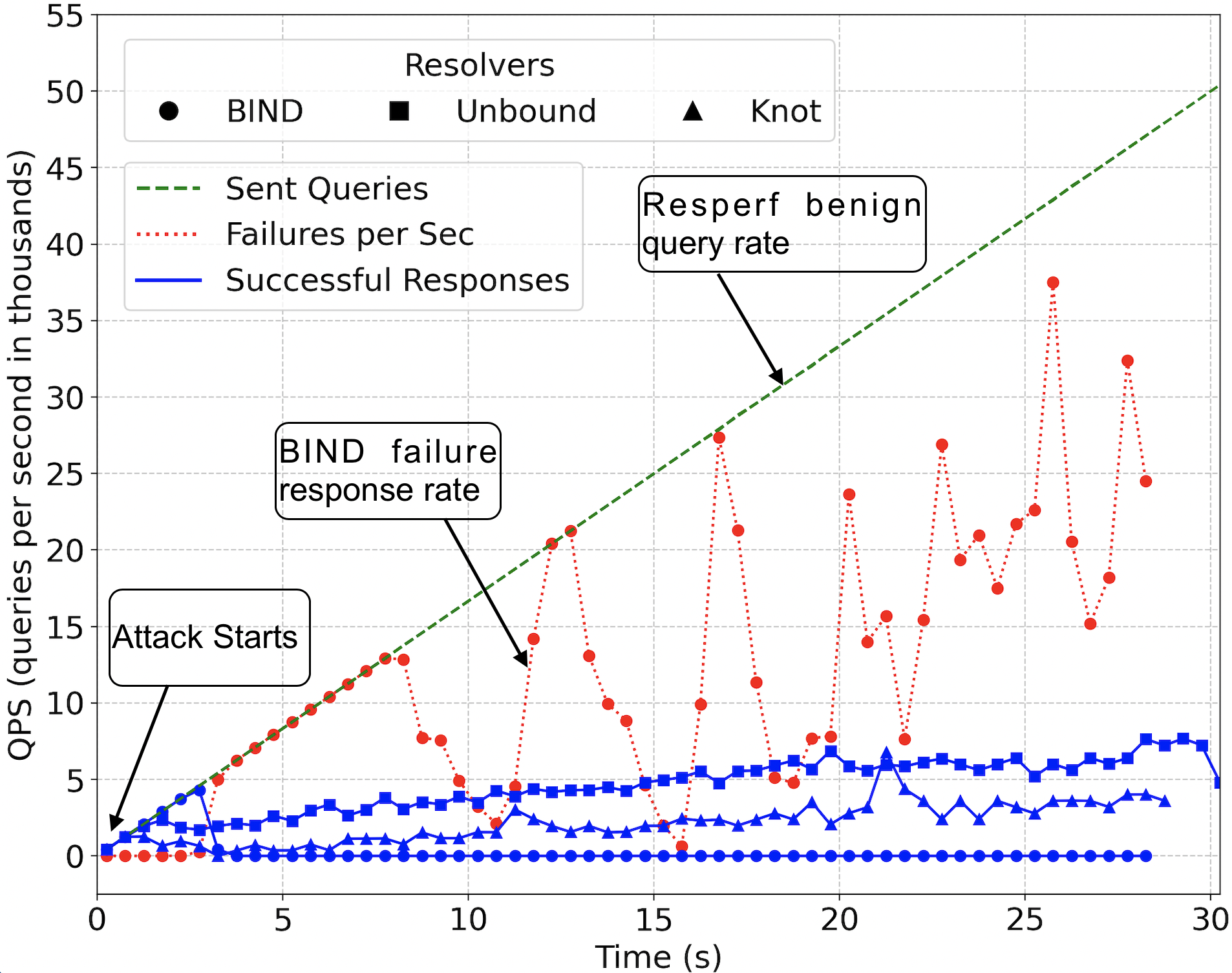} 
    \caption{Max benign response rates of resolvers under a constant 1,000 QPS rate bait-and-switch attack.}
    \label{fig:max_detail}
\end{figure}
}
\hideforifip{\begin{figure*}[ht]
    \centering
    \includegraphics[width=\textwidth,height=0.25\textheight]{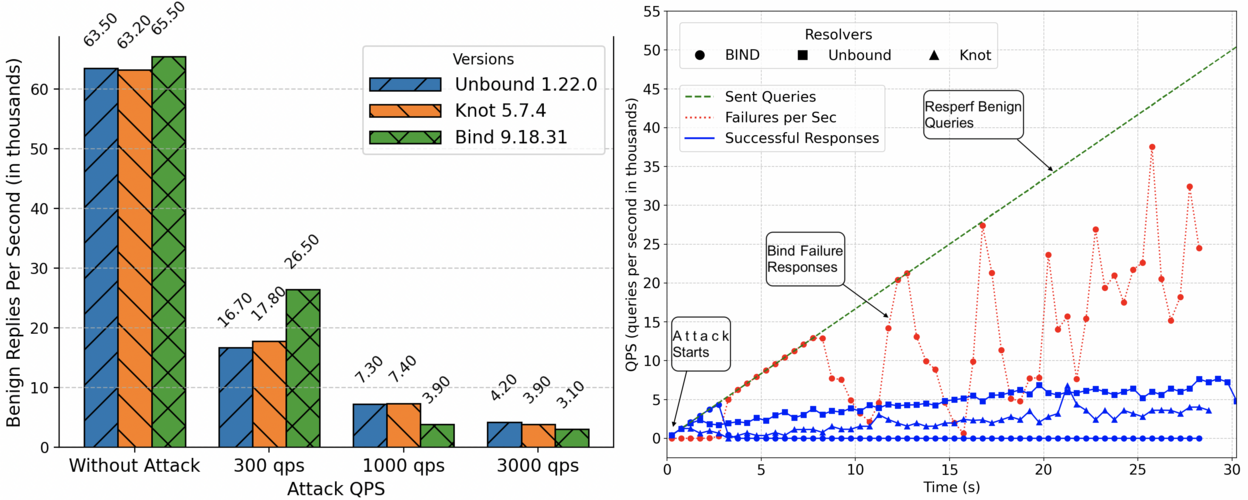} 
    \caption{a)Resolver Benign Max QPS Performance with Bait\&Switch Attack (excluding SERVERFAIL responses) under constant attack rates of 0, 300, 1,000, and 3,000 QPS. b) Detailed benign response rates of resolvers benign max QPS under a bait-and-switch attack at constant 1,000 QPS. The attack starts at 0s, marked by the black arrow.\rca{These figures were separete. They are merged to save space.}}
    \label{fig:max_qps_detail}
\end{figure*}}

\hideforifip{\begin{figure*}[ht]
    \centering
    \includegraphics[width=\textwidth]{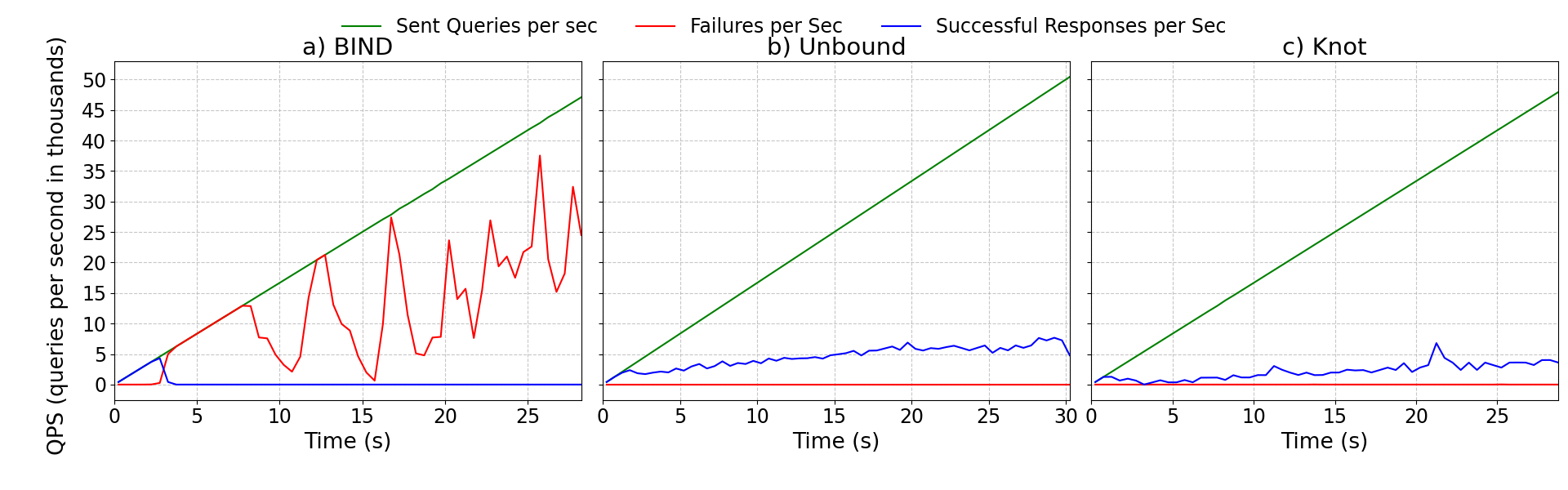} 
    \caption{Detailed benign response rates of resolvers benign max QPS under a bait-and-switch attack at constant 1,000 QPS. The attack starts at 0s, marked by the black arrow.}
    \label{fig:max_qps_detail}
\end{figure*}}

\hideforifip{
\subsection{Average Throughput Tests Under a Longer Attack}
Maximum benign QPS tests reveal the resolver’s peak capacity under attack but only as transient values. To assess its average performance in a sustained attack, we conducted a second Resperf experiment using the same zone and query files as the max-QPS test. Each resolver was initially started without any attack, and the benign QPS was gradually increased to 50,000, filling the resolver's cache with benign domains. Benign queries continued at this rate for 10 seconds, and all responses were served from the cache. The attack then ran at 1K QPS for 130 seconds, revealing the resolver's behavior under sustained load.
\hideforifip{
We calculated the average benign QPS after the attack fully took effect (following the first significant QPS drop). The average throughput test results for BIND, Unbound, and Knot are presented in Figure-\ref{fig:avg_performance}. During the attack, BIND's benign traffic rapidly drops to 0\%, leaving it incapable of answering benign queries for the remainder of the attack. Unbound's benign QPS gradually declines to zero over approximately 30 seconds, after which its cache is completely flushed. Beyond this point, Unbound experiences a steady 80\% throughput drop for the duration of the attack. For Knot, the attack takes effect immediately; however, the resolver's throughput fluctuates more than Unbound's, with an average throughput drop of 73\%. This test represents a more realistic attack scenario, as resolver caches are typically filled with benign domains when an attack begins.}

\showforifip{We calculated the average benign QPS after the attack fully took effect, following the first significant QPS drop. Figure-\ref{fig:avg_performance} (a,b,c) presents the results for BIND, Unbound, and Knot. During the attack, BIND’s benign traffic drops to 0\%, rendering it unable to respond to queries. Unbound's QPS declines to zero over 30 seconds as its cache is flushed, then stabilizes at an 80\% throughput drop. Knot is affected immediately, with throughput fluctuating but averaging a 73\% drop. This test represents a more realistic attack scenario, as resolver caches are typically filled with benign domains when an attack begins.}


\hideforifip{\begin{figure*}[ht] 
    \centering
    \includegraphics[width=\textwidth,height=0.20\textheight]{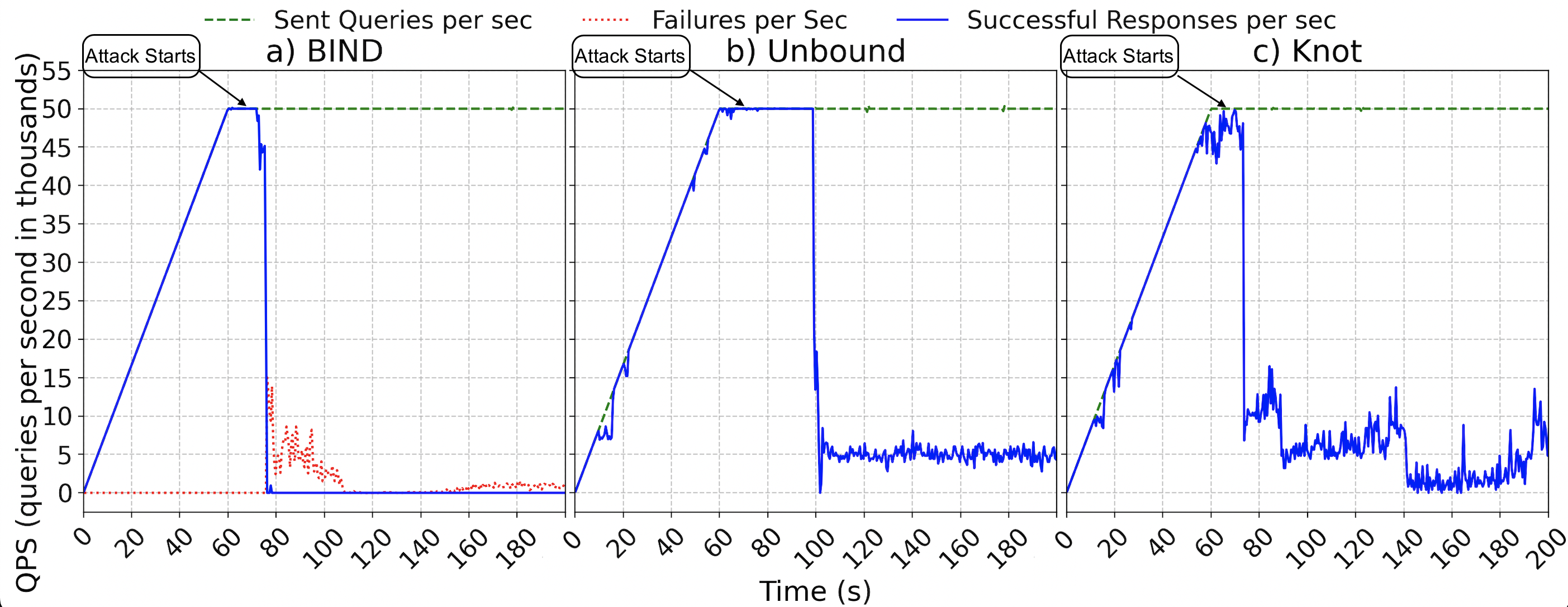} 
    \caption{Resolver benign average QPS under a bait-and-switch attack at constant 1,000 QPS. The attack starts at 70s.}
    \label{fig:avg_performance}
\end{figure*}}

\hideforifip{
\begin{figure*}[ht]
   \centering
    \includegraphics[width=\textwidth]{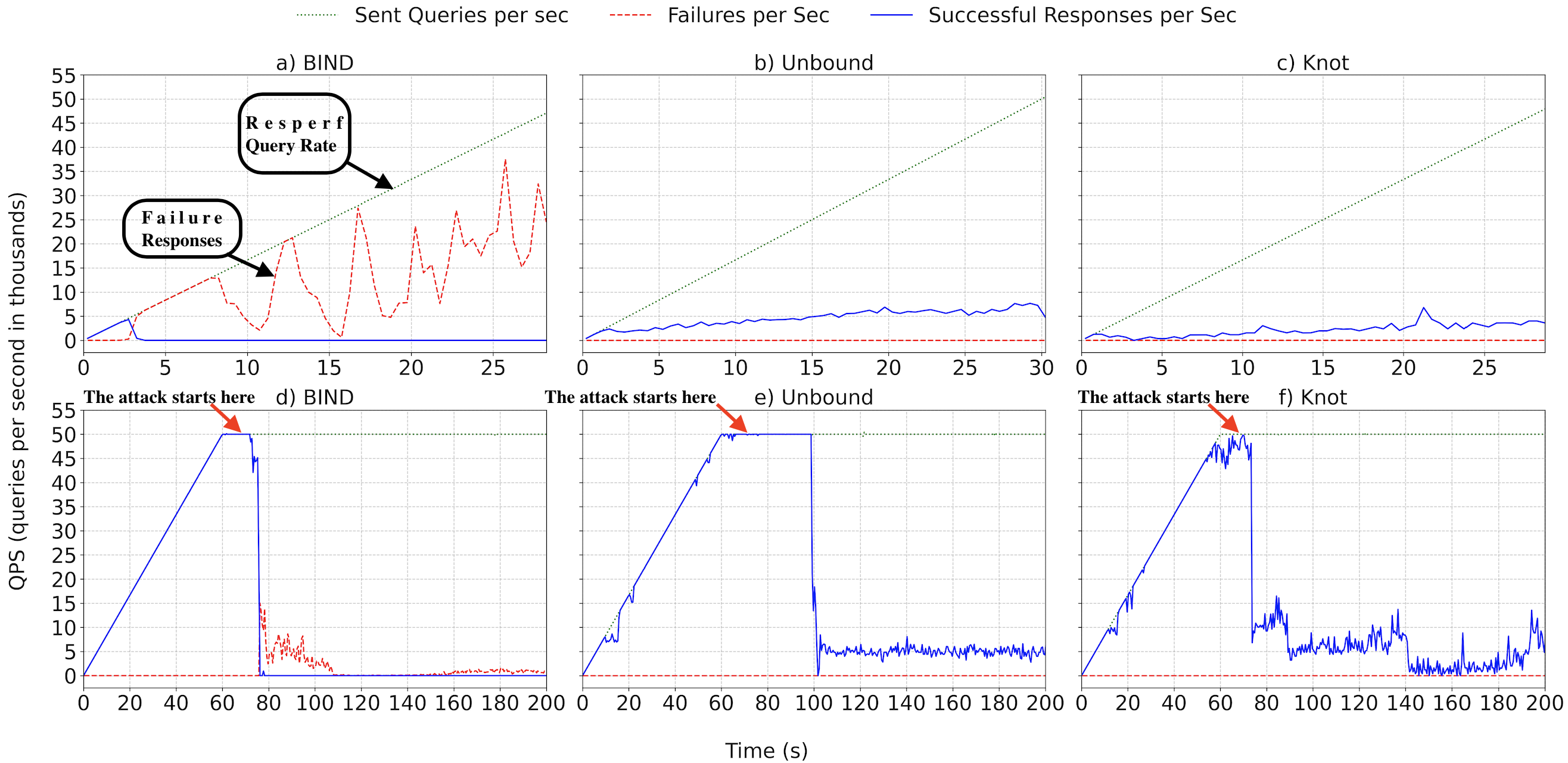} 
    \caption{The top half of the figure (a, b, and c) shows the detailed response rates of resolvers during the maximum throughput test. The steady line with dots represents Resperf's benign query rate, while dashed lines indicate server failures, which are significant only for BIND. The bottom half (d, e, and f) displays the resolver's benign average QPS under a bait-and-switch attack at a constant 1,000 QPS. The attack starts at 70s and lasts for 130s, marked by the red arrow.}
    \label{fig:avg_max_combined_performance}
\end{figure*}

}

}

\hideforifip{

\subsection{Re-discovering Existing DNS Vulnerabilities\ignore{ with LAPRAD}}

LAPRAD assisted also in the re-discovery of Keytrap \cite{heftrig_harder_2024} and CacheFlush(2024) \cite{afek2024flushing} attacks.

We first confirmed GPT's knowledge cutoff as October 2023 by testing its awareness of publicly available information released afterward. After that, we checked whether it recognized the KeyTrap attack \cite{heftrig_harder_2024} (published February 2024) or the CacheFlush paper \cite{afek2024flushing} (released August 2024). In both cases, it showed no awareness, confirming its knowledge cutoff. Due to space limitations, related prompts are omitted and can be found in \cite{all_gpt_conversations}.

}

\ignore{
\rca{Added this paragraph and the table-II}
}
\ignore{
\textcolor{blue}
}
Table-\ref{tab:open_resolvers} presents the caching behavior of open resolvers for the newly discovered \textbf{\textit{SigCacheFlush}} vulnerabilities. First, we queried the resolver for large DNSSEC records of a domain under our control, then terminated the domain's authoritative server. If the resolver continued responding to queries for this domain as before, it indicated successful caching; otherwise, it was merely relaying responses. Each resolver was tested only once to minimize disruption to legitimate services.
\vspace{-0.3cm}
\begin{table}[htbp] 
    \begin{tabularx}{\columnwidth}{|l||X|X|X|}
    \hline
    \textbf{Open Resolver} & \textbf{B\&S} & \textbf{MK} & \textbf{ANY} \\ \hline
    \hline
    Google Public DNS (8.8.8.8) & \checkmark & \checkmark & \checkmark \\ \hline
    Cloudflare DNS (1.1.1.1) & \checkmark & \checkmark & \textcolor{red}{\texttimes} \\ \hline
    Quad9 (9.9.9.9) & \checkmark & \checkmark & \checkmark \\ \hline
    OpenDNS -Cisco (208.67.222.222) & \checkmark & \checkmark & \textcolor{red}{\texttimes} \\ \hline
    \end{tabularx}
    \caption{Comparison of Open Resolvers for Bait and Switch (B\&S), Multiple RSA Keys (MK), and ANY-Type SigCacheFlush Attacks. (\checkmark) indicates that the resolver supports (or is vulnerable to) the feature; (\textcolor{red}{\texttimes}) means the opposite. }
    \label{tab:open_resolvers}
\end{table}
\vspace{-0.5cm}

\vspace{-0.4cm}
\section{Discussions and Future Work}
\hideforifip{\textit{\textbf{Combining Fine Protocol Nuances for Attack Discovery}}:
Discovering complex attacks, such as bait-and-switch, requires a deep understanding of DNSSEC features. For instance, RRsets can be signed with multiple keys, RRSIGs are constrained only by the maximum DNS packet size, and a single valid signature enables caching of the entire RRSet. The LLM effectively combined these nuances to formulate the investigated attack idea.

\textit{\textbf{Automating the LLM-Assisted Attack Discovery Process}}:
}
Currently, human operators drive the attack investigation process by focusing on answers they believe might lead to new attacks. This approach limits the scope of attack discovery, as it is influenced by the operator’s intuition. Despite this limitation, we demonstrated that a viable attack could be achieved using a small number of general questions (2-7), with the only bias being the selection of target answers. This underscores the potential for further automation in the attack discovery process. 

In future work, we aim to eliminate human involvement in the attack discovery loop by automating Step 1 of LAPRAD using ReACT \cite{yao2022react} and multi-LLM agents. This will enable dynamic decision-making and iterative refinement during attack idea generation and RFC-based fact-checking.

\hideforifip{
\begin{figure*}[ht]
    \centering
    \includegraphics[width=\textwidth]{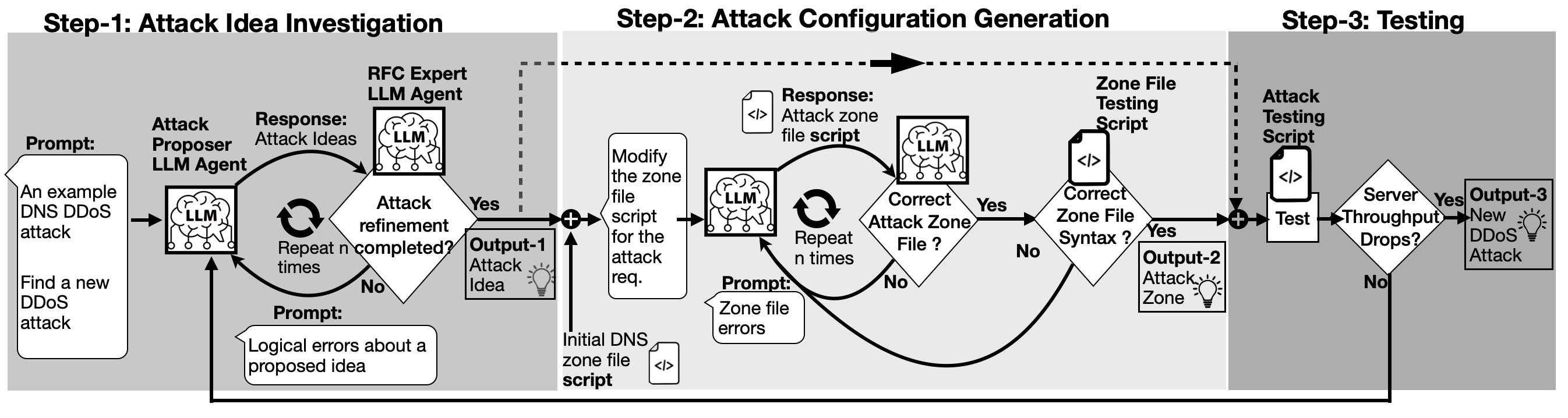} 
    \caption{Multi-Agent-Based Automatic Network Vulnerability Discovery Framework}
    \label{fig:future_design}
\end{figure*}
}

\section{Conclusion}
This paper introduced LAPRAD methodology which has shown the effectiveness of LLMs in identifying, analyzing, and addressing vulnerabilities in the DNSSEC protocol. Given a DNS attack targeting the resolver cache, LAPRAD led to the proposal of three new variants of the initial example attack, each exploiting completely different protocol features than the original. By leveraging detailed protocol features and adapting to constraints, LAPRAD successfully generated sophisticated attack ideas, such as the bait-and-switch and KeyTrap attacks.

We identified previously unknown DNSSEC-related vulnerabilities and confirmed our approach by testing LLMs on existing vulnerabilities not included in their training set. The methodology we outlined can be applied to any network protocol or software system with well-documented standards, such as protocol RFCs.

Our findings demonstrate that LLM-assisted network vulnerability investigation is a promising foundation for further automation. Moving forward, we aim to create a highly automated LAPRAD framework that minimizes human involvement while maximizing the discovery of vulnerabilities. (An extended version of this study is available at \cite{all_gpt_conversations}.)
\showforifip{
\section*{Acknowledgments}
We thank George Varghese (UCLA) for helpful discussions and Shoham Danino for assistance with the testing environment. We thank Petr Špaček (ISC), Yorgos Thessalonikefs (Unbound), Otto Moerbeek (PowerDNS), and other DNS-OARC Mattermost members for their role in responsible disclosure and helpful discussions. This work was supported in part by a grant from the Sunday Group to The UCLA Connection Lab, as well as by ISF Grant No. 1527/23, a grant from the Blavatnik Interdisciplinary Cyber Research Center (ICRC) at Tel Aviv University, and the Blavatnik Family Fund.
}

\bibliographystyle{IEEEtran}
\bibliography{citations}
\hideforifip{\section{Ethics considerations} 
\rca{this whole section and the appendix will go away. They consider appendix in the 9 page limit}
This research investigates vulnerabilities in the DNS protocol. All experiments involving the discovered vulnerabilities were conducted in a controlled laboratory environment, isolated from regular Internet traffic.

For the vulnerabilities identified in this study, we are currently in the responsible disclosure process. We have informed vendors about these vulnerabilities through the confidential OARC-Mattermost channel and created a ticket on the vendors' GitLab channels. We are collaborating with vendors on CVE assignments and fixes. Any fixes will be issued before the paper's publication date, and we are coordinating with developers to schedule them accordingly.

We performed open resolver tests only once for Google Public DNS, Cloudflare DNS, Quad9, and OpenDNS-Cisco to avoid any disruption to these services. We also plan to inform these open resolver vendors in the near future.

\section{Open science} 
To comply with the open science policy, we are publishing an artifact alongside this paper that includes all data, GPT conversations, code, and configurations required for reproducibility. This ensures our results can be independently verified and promotes transparency and collaboration in research.
\section*{Availability}
Currently, our GPT conversations are shared. In addition, we will provide the testing Docker images to reproduce the attack tests described in this paper.
\appendix
\section{Appendix A}
\label{appendix:A}
----------------------------------------------------------
\begin{figure*}[ht]
    \centering
    \includegraphics[width=\textwidth]{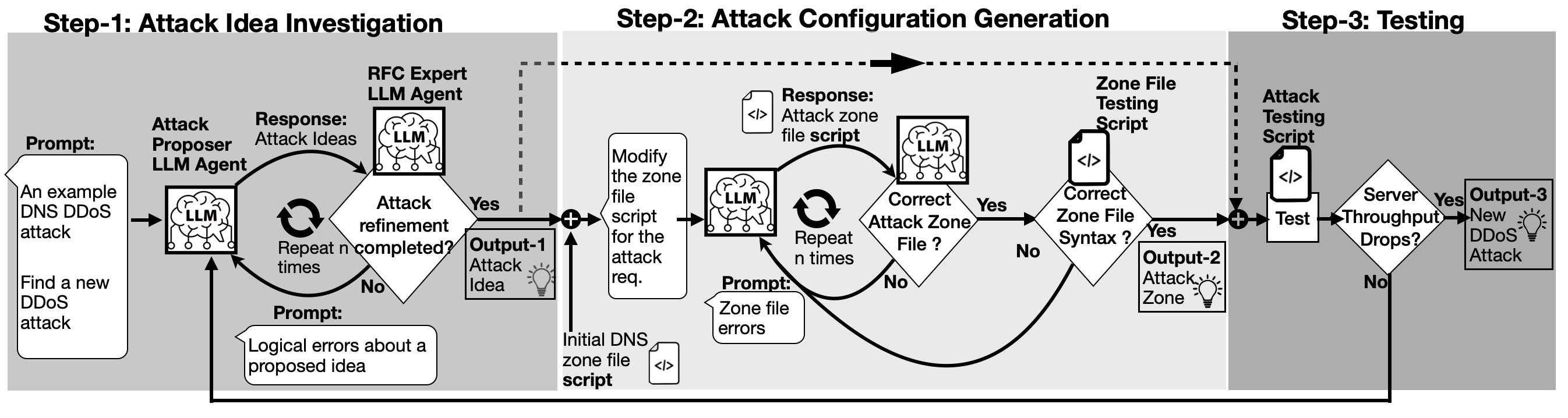} 
    \caption{Multi-Agent-Based Automatic Network Vulnerability Discovery Framework}
    \label{fig:future_design}
\end{figure*}

\gptpromtbox{mitigation_prompt}{
An attack summary (Output-1, obtained in \textbf{\textit{Step-1}}))\\

What would be the most logical mitigation for this conceptual attack that does not affect the other parts of the DNS system?

\gptanswerbox{mitigation_answer}{
\textbf{\textit{Bait \& Switch Cache Flush}}\\
Size Limits on RRSIG Records
\hl{A resolver could impose an upper bound on RRSIG size} ...
RSA-8192 might produce a 1–2 kilobyte signature...
Thus, a threshold of, say,\hl{ a few kilobytes (2k–4k bytes) might be enough to handle current mainstream algorithms comfortably}.

\textbf{\textit{Multi RSA 4096 Key Cache Flush}}\\
Local Policy on RRset Size
\hl{Most real-world DNSSEC-signed zones have modest DNSKEY RRsets (e.g., 1–4 DNSKEYs). Setting a reasonable upper limit (e.g., 20 DNSKEY records or 32 KB total) will hardly impact legitimate domains.}

\textbf{\textit{ANY Type Cache Flush}} \\ 
\hl{A record cap (e.g., “Cache no more than 100 total RRs for a given name”)} prevents any single domain from displacing too much data in the cache.
... evil.example, returns a massive 500-record DNS response (e.g., via an ANY or DNSSEC query). When \hl{the resolver receives that response, it stores only the first 100 records and discards the remainder}.
}
}
}
\end{document}